\documentclass[12pt]{article}
\usepackage{a4wide}
\usepackage{amssymb}
\usepackage{amsmath}
\usepackage{graphicx}
\usepackage{mathdots}
\usepackage{slashed}
\usepackage{verbatim}
\usepackage{xcolor}
\usepackage{cancel}

\usepackage{cite}

\usepackage{hyperref}

\usepackage{IEEEtrantools}

\def\makeatletter{\catcode`\@=11}
\makeatletter
\def\mathbox#1{\hbox{$\m@th#1$}}%
\def\math@ccstyles#1#2#3#4#5#6#7{{\leavevmode
      \setbox0\mathbox{#6#7}%
      \setbox2\mathbox{#4#5}%
      \dimen@ #3%
      \baselineskip\z@\lineskiplimit#1\lineskip\z@
      \vbox{\ialign{##\crcr
             \hfil \kern #2\box2 \hfil\crcr
             \noalign{\kern\dimen@}%
             \hfil\box0\hfil\crcr}}}}
\def\mathaccstyles{\math@ccstyles\maxdimen}
\def\maththroughstyles{\math@ccstyles{-\maxdimen}}
\def\unity%
 {\maththroughstyles{.45\ht0}\z@\displaystyle {\mathchar"006C}\displaystyle 1}

\numberwithin{equation}{section}
 
\begin{document}

\mbox{}
\vspace{0truecm}
\linespread{1.1}


\centerline{\Large \bf RG Flows and Stability in Defect Field Theories }




\vspace{.4cm}

 \centerline{\LARGE \bf }

\vspace{1.5truecm}

\centerline{
    { \bf I.Carre\~no Bolla${}^{a,b}$} \footnote{ignaciocarbolla@gmail.com}
    { \bf D. Rodriguez-Gomez${}^{a,b}$} \footnote{d.rodriguez.gomez@uniovi.es}
   {\bf and}
    { \bf J. G. Russo ${}^{c,d}$} \footnote{jorge.russo@icrea.cat}}

\vspace{1cm}
\centerline{{\it ${}^a$ Department of Physics, Universidad de Oviedo}} \centerline{{\it C/ Federico Garc\'ia Lorca  18, 33007  Oviedo, Spain}}
\medskip
\centerline{{\it ${}^b$  Instituto Universitario de Ciencias y Tecnolog\'ias Espaciales de Asturias (ICTEA)}}\centerline{{\it C/~de la Independencia 13, 33004 Oviedo, Spain.}}
\medskip
\centerline{{\it ${}^c$ Instituci\'o Catalana de Recerca i Estudis Avan\c{c}ats (ICREA)}} \centerline{{\it Pg.~Lluis Companys, 23, 08010 Barcelona, Spain}}
\medskip
\centerline{{\it ${}^d$ Departament de F\' \i sica Cu\' antica i Astrof\'\i sica and Institut de Ci\`encies del Cosmos}} \centerline{{\it Universitat de Barcelona, Mart\'i Franqu\`es, 1, 08028
Barcelona, Spain }}
\vspace{1cm}
\setcounter{footnote}{0}

\centerline{\bf ABSTRACT}
\medskip 

We investigate defects in scalar field theories in four and six dimensions in a double-scaling (semiclassical) limit,
where bulk loops are suppressed and quantum effects come from the defect coupling. We compute $\beta $-functions up to four loops and find that fixed points satisfy dimensional disentanglement --{\it i.e.} their dependence on the space dimension is factorized from the coupling dependence-- and discuss some physical implications.
We also give an alternative derivation of the $\beta$ functions by computing  systematic logarithmic corrections to the Coulomb potential.
In this natural scheme, $\beta $ functions turn out to be a gradient of a `Hamiltonian' function ${\cal H}$.
We also obtain closed  formulas for the dimension of
scalar operators
and show that instabilities do not occur for  potentials bounded from below. The same  formulas are reproduced using  Rigid Holography.

\noindent 

\newpage

\tableofcontents

\section{Introduction}

A Quantum Field Theory (QFT) generically contains extended, non-local, operators supported on lower-dimensional manifolds. It is fair to say that these have been, at least comparatively, much less studied than the very familiar local operators. Yet, they can provide interesting new insights into QFT from
 their Renormalization Group (RG) flows and from associated (generalized) symmetries.  At present, defects and boundaries are being intensively studied from various points of view  (see \cite{Lauria:2020emq,Bianchi:2021snj,Gimenez-Grau:2022czc,Gimenez-Grau:2022ebb, Bianchi:2022sbz, Popov:2022nfq, Cuomo:2022xgw,Cuomo:2021kfm,Rodriguez-Gomez:2022gbz,Rodriguez-Gomez:2022gif,Cuomo:2021rkm,Shachar:2022fqk,Aharony:2022ntz,Beccaria:2017rbe,Herzog:2019bom,Herzog:2020lel,Giombi:2020amn,Giombi:2020rmc,Giombi:2021cnr,Drukker:2020swu,Herzog:2022jlx,Giombi:2022vnz,Pannell:2023pwz,Drukker:2022pxk} for a list of recent developments).

An approach that has proven to be very useful in many instances is to search corners in the coupling parameter space in which to perform a controlled perturbative approximation. The semiclassical approximation itself is an example of this paradigm. Other examples include the large $N$ approximation or the study of large spin sectors. A novel method introduced recently consists in the study of sectors of operators with large charge under a global symmetry (see \cite{Gaume:2020bmp} for a review and references). 
The method used in this paper is similar. This has been considered in \cite{Rodriguez-Gomez:2022gbz,Rodriguez-Gomez:2022gif}
to study different aspects of (flat) defects in scalar field theories in $d=4-\epsilon$ and $d=6-\epsilon$ dimensions, by assuming a  scaling limit of the couplings, where the defect couplings are large and the bulk couplings are small. As a result, quantum effects in the bulk vanish, while the defect still induces  non-trivial  quantum dynamics. In particular one can study the RG flow of the defect couplings and find interesting phenomena such as fixed point creation/annihilation. The results in \cite{Rodriguez-Gomez:2022gif} show, quite surprisingly, that the  position of such fixed points is set by the one-loop approximation up to an overall scale  that solely depends on $\epsilon$. This separation of the dimension and coupling dependence  is in general unexpected and it has been dubbed \textit{Dimensional Disentanglement} (DD) in \cite{Rodriguez-Gomez:2022gif}. Additionally, the position of the fixed points can be dialed by tuning the  bulk couplings, which act as knobs that can be adjusted.

In this paper we set out to study in more depth these aspects for flat defects in scalar field theories both in $d=4-\epsilon$ dimensions (where the defect is a line of codimension $d_T=3-\epsilon$) and in $d=6-\epsilon$ dimensions (where the defect is a surface of codimension $d_T=4-\epsilon$) . In particular, we extend the explicit two-loop computation of the defect $\beta$ functions in \cite{Rodriguez-Gomez:2022gif} to four loops. This supports a conjecture that DD is actually a universal property holding for any theory in the double-scaling limit. 

It was noticed  in \cite{Rodriguez-Gomez:2022gif} that  the two-loop $\beta$ functions of the defect couplings
are the gradient of a function ${\cal H}$, where $\exp({\cal H})$ matches the VEV of the circular defect. This has been proposed to reflect  monotonic properties of the defect RG flow in \cite{Cuomo:2021rkm}. Similar observations  have been recently made in \cite{Shachar:2022fqk} for the 6d case, considering now a spherical two-dimensional defect.

Starting with three loops, the $\beta$ functions contain scheme-dependent corrections. In the scheme of section 2 based on dimensional regularization, we find that the $\beta $-functions are no longer a gradient beyond two loops. The freedom left by the choice of scheme 
raises the question of whether there could be a scheme such that the $\beta$-functions are still a gradient of a function (as conjectured in \cite{Rodriguez-Gomez:2022gif}). This question is answered positively in section 3: an alternative calculation of the $\beta$ function using the dressed Coulomb potential gives $\beta_i=2\,c\,\partial_i\mathcal{H}$ up to four loop orders. We explicitly provide a formula for  $\mathcal{H}$ for any 4d or 6d scalar field theory with general marginal potentials.

Using our results for the $\beta$ functions, we  construct theories in which, for $\epsilon=0$, both bulk and defect couplings are at a fixed point. These models thus define defect Conformal Field Theories (dCFT's). Given a dCFT, a problem of interest is to see if the theory may suffer from instabilities due to the presence of dangerously irrelevant operators.\footnote{A dangerously irrelevant operator is an operator that is naively irrelevant but approaches marginality at certain critical values of the parameters of the theory. In a CFT they typically signal the presence of a nearby fixed point and hint to instabilities.} Following  \cite{Aharony:2022ntz}, we study these possible instabilities in our theories, finding that they are absent provided that the potential is bounded from below. We also study a fermion-scalar theory with a Yukawa interaction in 4d, which perturbatively defines a dCFT, in search for such instabilities, finding also that they are absent.

When $\epsilon= 0$, the double-scaling limit freezes the running of bulk couplings and the bulk theory becomes conformally invariant. In appendix C we  make use of this property to engineer a setup suitable for holographic methods. As $\mathbb{R}^d$ is conformal to $\mathbb{H}^{d_T-1}\times \mathbb{S}^{d_T-1}$, the theory can be directly put in $\mathbb{H}^{d_T-1}\times \mathbb{S}^{d_T-1}$.
Then the boundary of the $\mathbb{H}^{d_T-1}$ is identified with the defect. This is similar in spirit   to \textit{rigid holography} \cite{Aharony:2015zea} (for further developments along these lines, see \textit{e.g.} \cite{Paulos:2016fap,Beccaria:2017rbe,Carmi:2018qzm,Herzog:2019bom,Herzog:2020lel,Giombi:2020amn,Giombi:2020rmc,Giombi:2021cnr}). In our approach, we make use of this idea to compute defect $\beta$ functions, finding a precise agreement with the field theory results.

\section{Defects in scalar field theories and dimensional disentanglement}\label{sec:Defectsinscalarfieldtheories}

We consider a general theory with $N$ scalar fields in $d=4-\epsilon$, $d=6-\epsilon$ dimensions. Denoting the fields $\Phi_i$, with $i=1,\cdots , N$, we consider the following  action in Euclidean signature,
\begin{equation}
S=\int d^dx\, \left(\frac{1}{2}(\partial\Phi_i)^2+V(\Phi_i)\right)\,,
\end{equation}
where $V$ is a generic homogeneous polynomial in the $\Phi_i$'s of strict degree $n$, with couplings $\hat g_\alpha $. In $d=4-\epsilon$ dimensions $n=4$, while in $d=6-\epsilon$ dimensions $n=3$. 
That is, $V$ is of the form $V=\sum_\alpha \hat g_{\alpha}\Phi_i\Phi_j\Phi_k\Phi_l$, with $\alpha=[i,j,k,l]$ in $d=4-\epsilon$, and 
$V=\sum_\alpha \hat g_{\alpha}\Phi_i\Phi_j\Phi_k$, with $\alpha=[i,j,k]$ in $d=6-\epsilon$.

We now consider a trivial defect which is a line in $d=4-\epsilon$ dimensions and a surface in $d=6-\epsilon$ dimensions. Hence the dimension of the worldvolume is $1$ in $d=4-\epsilon$ dimensions and $2$ in $d=6-\epsilon$ dimensions, while the dimension of the transverse space is  $d_T=3$ in the 4d theory and $d_T=4$ in the 6d theory. In both cases the defect admits a (slightly relevant for $\epsilon\ne 0$) deformation by the $\Phi_i$'s. Thus, we are led to consider the defect theory with action

\begin{equation}
\label{trea}
S=\int d^dx\, \left(\frac{1}{2}(\partial\Phi_i)^2+V(\Phi_i)-h_i\,\Phi_i\,\delta_T\right)\, ,
\end{equation}
where $\delta_T$ denotes the Dirac delta function in the transverse space to the defect.
We are now interested in a particular scaling limit of both the defect and bulk couplings (bulk couplings  are collectively denoted by $\hat{g}_{\alpha}$). Specifically, we are interested in a situation where  the defect couplings are very large and the bulk couplings are small, keeping $\hat{g}_{\alpha} h_i^{n-2}$ fixed. In this limit, pure bulk loop corrections that do not involve $h_i$ couplings are suppressed, while quantum effects get organized in powers of this effective finite coupling $(\hat{g}_{\alpha} h_i^{n-2})$.

To implement this limit, one can formally introduce new variables as follows:
\begin{equation}
\label{eq:limit}
h_i=\hbar^{-\frac{1}{2}}\,\,\nu_i\,,\qquad \hat{g}_{\alpha}=\hbar^{\frac{n-2}{2}}\,g_{\alpha}\,, \qquad \Phi_i=\hbar^{-\frac{1}{2}}\,\phi_i\ .
\end{equation}
This gives
\begin{equation}
\label{pres}
S=\frac{1}{\hbar }\,S_{\rm eff}\,,\qquad S_{\rm eff}=\int d^dx\, \left(\frac{1}{2}(\partial\phi_i)^2+V(\phi_i)-\nu_i\,\phi_i\,\delta_T\right)\,.
\end{equation}
Thus, we see that a semiclassical limit exists where $\hbar\rightarrow 0$ while $\nu_i$, $g_{\alpha}$ are  fixed. 
In the following we will take this limit and explore its consequences, specializing to flat defects.

\subsection{Solving the saddle-point equation in perturbation theory}

In the double-scaling limit introduced above, there is a semiclassical expansion for $S_{\rm eff}$. The corresponding equations of motion are
\begin{equation}
\partial^2\phi_i-V_i=-\nu_i\,\delta_T\,,
\end{equation}
where the subscript in $V$ means derivative with respect to $\phi_i$. 

We will solve these equations in perturbation theory in the bulk couplings, extending a calculation done in \cite{Rodriguez-Gomez:2022gif} to higher orders. To that matter we write $\phi_i=\phi_i^{(0)}+\phi_i^{(1)}+\phi_i^{(2)}+\phi_i^{(3)}+\phi_i^{(4)}+\phi_i^{(5)}\cdots$. The equation becomes

\begin{eqnarray}
\nonumber && \partial^2\phi^{(0)}_i+\partial^2\phi^{(1)}_i+\partial^2\phi^{(2)}_i+\partial^2\phi^{(3)}_i+\partial^2\phi^{(4)}_i+\partial^2\phi^{(5)}_i+\cdots \\ \nonumber &&-V_i-V_{ij}\phi_j^{(1)}-\Big(V_{ij}\phi_j^{(2)}+\frac{1}{2}\,V_{ijk}\,\phi_j^{(1)}\,\phi_k^{(1)}\Big)-\Big( V_{ij}\,\phi^{(3)}_j+V_{ijk}\,\phi^{(1)}_j\,\phi^{(2)}_k+\frac{1}{6}\,V_{ijkl}\,\phi^{(1)}_j\,\phi^{(1)}_k\,\phi^{(1)}_l\Big)-\\ \nonumber &&
\Big( V_{ij}\,\phi^{(4)}_j+\frac{1}{2}\,V_{ijk}\,\phi^{(2)}_j\,\phi^{(2)}_k+V_{ijk}\,\phi^{(1)}_j\,\phi^{(3)}_k+\frac{1}{2}\,V_{ijkl}\,\phi^{(1)}_j\,\phi^{(1)}_k\,\phi^{(2)}_l+\frac{1}{24}\,V_{ijklm}\phi^{(1)}_j\,\phi^{(1)}_k\,\phi^{(1)}_l\,\phi^{(1)}_m\Big)-\cdots\\  && =-\nu_i\,\delta_T\,.
\nonumber
\end{eqnarray}
Here $V$ and its derivatives are evaluated at $\phi_i^{(0)}$. 
We can now solve order by order.

\medskip

\noindent \textbf{Order 0}: The equation is

\begin{equation}
\partial^2\phi^{(0)}_i=-\nu_i\,\delta_T\qquad \implies \qquad \phi_i^{(0)}=\nu_i\,\int d^d z_1\,G(x-z_1)\,\delta_T(z_1)\,.
\end{equation}
It will turn out convenient to introduce the function

\begin{equation}
\phi=\int d^dz_1\,G(x-z_1)\,\delta_T(z_1)\,.
\end{equation}
Using that $\phi_i^{(0)}=\nu_i\,\phi$, and the fact  that $V$ is a homogeneous degree $n$ function, we have the identity,

\begin{equation}
\label{eq:Vs}
V_{i_1\cdots i_m}(\phi_i^{(0)})=V_{i_1\cdots i_m}(\nu_i)\,\phi^{n-m}\,.
\end{equation}
\medskip

\noindent \textbf{Order 1}: At this order we have

\begin{equation}
\partial^2\phi^{(1)}_i=V_i \qquad \implies  \qquad \phi_i^{(1)}=-\int d^d z_1\,G(x-z_1)\,V_i(z_1)\,.
\end{equation}
Therefore, using \eqref{eq:Vs},

\begin{equation}
\phi_i^{(1)}=-V_i\,I_1\qquad  I_1= \int d^dz_1\,G(x-z_1)\,\phi(z_1)^{n-1}\,,
\end{equation}
where $V_{i_1\cdots i_m}$ refers now to $V_{i_1\cdots i_m}(\nu_i)$.
To lighten the notation, let us define:
\begin{equation}
    \hat G_r(x,y)\equiv G(x-y)\phi(y)^r\, ,
\end{equation}
so that
\begin{equation}
 I_1= \int d^dz_1\, \hat G_{n-1}(x,z_1)\,.
\end{equation}

\medskip

\noindent \textbf{Order 2}: The equation is now

\begin{equation}
\partial^2\phi^{(2)}_i=V_{ij}\phi_j^{(1)}\qquad \implies  \qquad \phi_i^{(2)}=-\int d^d z_1\,G(z_1-y)\,V_{ij}\phi_j^{(1)}(y)\,.
\end{equation}
Hence

\begin{equation}
\phi_i^{(2)}=V_j\,V_{ij}\,I_2\,,\qquad I_2=\int d^d z_1\, d^d z_2\,\hat G_{n-2}(x,z_1)\,\,\hat G_{n-1}(z_1,z_2)\,,
\end{equation}
where we have used \eqref{eq:Vs} to write the result in terms of $V_{i_1\cdots i_m}=V_{i_1\cdots i_m}(\nu_i)$.

\medskip

\noindent \textbf{Order 3}: The equation is

\begin{equation}
\partial^2\phi^{(3)}_i=\Big(V_{ij}\phi_j^{(2)}+\frac{1}{2}\,V_{ijk}\,\phi_j^{(1)}\,\phi_k^{(1)}\Big)\ .
\end{equation}
Therefore
\begin{equation}
 \phi_i^{(3)}=-\int d^d z_1\,G(x-z_1)\,\Big(V_{ij}\phi_j^{(2)}+\frac{1}{2}\,V_{ijk}\,\phi_j^{(1)}\,\phi_k^{(1)}\Big)\, ,
\end{equation}
and
\begin{equation}
\phi_i^{(3)}=-V_{ij}V_{jk}\,V_k \, I_3^{(1)}-\frac{1}{2}\,V_{ijk}\,V_j\,V_k \, I_3^{(2)}\,.
\end{equation}
Note that, once again, we have used \eqref{eq:Vs} to write the result in terms of $V_{i_1\cdots i_m}=V_{i_1\cdots i_m}(\nu_i)$. In addition

\begin{equation}
I_3^{(1)}=\int d^d z_1\, d^d z_2\ d^d z_3 \ \hat G_{n-2}(x,z_1)\,\hat G_{n-2}(z_1,z_2)\,\,\hat G_{n-1}(z_2,z_3)\,,
\nonumber
\end{equation}
\begin{equation}
I_3^{(2)}=\int d^d z_1\, d^d z_2\ d^d z_3 \ \hat G_{n-3}(x,z_1)\,\hat G_{n-1}(z_1,z_2)\,\hat G_{n-1}(z_1,z_3)\,\,.
\nonumber
\end{equation}

\medskip

\noindent \textbf{Order 4}: The equation is

\begin{equation}
\partial^2\phi^{(4)}_i=\Big( V_{ij}\,\phi^{(3)}_j+V_{ijk}\,\phi^{(1)}_j\,\phi^{(2)}_k+\frac{1}{6}\,V_{ijkl}\,\phi^{(1)}_j\,\phi^{(1)}_k\,\phi^{(1)}_l\Big)\,.
\end{equation}
Hence

\begin{equation}
 \phi^{(4)}_i=V_{ij}\,V_{jk}V_{kl}\,V_l\,I_4^{(1)}+\frac{1}{2}\, V_{ij}\,V_{jkl}\,V_k\,V_l\,I_4^{(2)} +V_{ijk}\,V_j\,V_{kl}V_l\,I_4^{(3)} +\frac{1}{6}\,V_{ijkl}\,V_j\,V_k\,V_l\,I_4^{(4)}\,,
\end{equation}
where $V_{i_1\cdots i_m}=V_{i_1\cdots i_m}(\nu_i)$ and

\begin{equation}
\hspace{-2cm} I_4^{(1)}= \int d^d z_1\,  d^d z_2\, d^d z_3\, d^d z_4 \ \hat G_{n-2}(x,z_1)\,\hat G_{n-2}(z_1,z_2)\,\hat G_{n-2}(z_2,z_3)\,\hat G_{n-1}(z_3,z_4)\, ,
\nonumber
\end{equation}
\begin{equation}
\hspace{-2cm} I_4^{(2)}= \int d^d z_1\, d^d z_2\, d^d z_3\ d^d z_4\ \hat G_{n-2}(x,z_1)\,\hat G_{n-3}(z_1,z_2)\,\hat G_{n-1}(z_2,z_3)\,\hat G_{n-1}(z_2,z_4)\, ,
\nonumber
\end{equation}
\begin{equation}
\hspace{-2cm} I_4^{(3)}=\int d^d z_1\ d^d z_2\ d^d z_3\ d^d z_4\, \hat G_{n-3}(x,z_1)\,\hat G_{n-1}(z_1,z_2)\,\hat G_{n-2}(z_1,z_3)\,\hat G_{n-1}(z_3,z_4)\, ,
\nonumber
\end{equation}
\begin{equation}
\hspace{-2cm} I_4^{(4)}=\int d^d z_1\ d^d z_2\ d^d z_3\ d^d z_4\,\hat G_{n-4}(x,z_1)\,\hat G_{n-1}(z_1,z_2)\,\,\hat G_{n-1}(z_1,z_3)\,\hat G_{n-1}(z_1,z_4)\,.
\nonumber
\end{equation}

\smallskip

We recall that $n=3$ in the 6d theory and $n=4$ in the 4d theory. The integral $I_{4}^{(4)}$ appears only in the 4d theory, since $V_{ijkl}$ vanishes in 6d. In fact, in the $n=3$ case all $V_{ijkl}$ appearing from order 5 on will vanish, simplifying the expressions of $\phi_i^{(m)}$.

Putting  everything together, we can write

\begin{equation}
\phi_i=-\int \frac{d^{d_T}\vec{p}^{\, T}}{(2\pi)^{d_T}}\,\frac{e^{i\vec{p}^{\, T}\cdot\vec{x}^T}}{(\vec{p}^{\, T})^2}\,\tilde{\phi}_i\,,
\end{equation}
with 
\begin{eqnarray}
&&\tilde{\phi}_i=-\nu_i+V_i\,\mathcal{I}_1-V_{ij}\,V_j\,\mathcal{I}_2+\Big(V_{ij}\,V_{jk}\,V_k\,\mathcal{I}_3^{(1)}+\frac{1}{2}\,V_{ijk}\,V_j\,V_k\,\mathcal{I}_3^{(2)}\Big)\\ \nonumber && -\Big(V_{ij}\,V_{jk}V_{kl}\,V_l\,\mathcal{I}_4^{(1)}+\frac{1}{2}\, V_{ij}\,V_{jkl}\,V_k\,V_l\,\mathcal{I}_4^{(2)} +V_{ijk}\,V_j\,V_{kl}V_l\,\mathcal{I}_4^{(3)} +\frac{1}{6}\,V_{ijkl}\,V_j\,V_k\,V_l\,\mathcal{I}_4^{(4)}\Big)\,.
\end{eqnarray}
The integrals are computed in appendix \ref{sect:integrals}. Focusing in the $d=4-\epsilon$ case, we finally obtain
\begin{eqnarray}
\tilde{\phi}_i &=& -\nu_i+V_i\,F_1\,|\vec{p}^{\, T}|^{2(d_T-3)}-V_{ij}\,V_j\,F_1\,F_{4-d_T}\,|\vec{p}^{\, T}|^{2(2d_T-6)}\\ \nonumber &&+\Big(V_{ij}\,V_{jk}\,V_k+\frac{1-5\epsilon}{1-3\epsilon}\,V_{ijk}\,V_j\,V_k\Big)\,F_1\,F_{4-d_T}\,F_{7-2d_T}\,|\vec{p}^{\, T}|^{2(3d_T-9)} \\ \nonumber &&  -\Big(V_{ij}\,V_{jk}V_{kl}\,V_l+V_{ij}\,V_{jkl}\,V_k\,V_l\,\frac{1-5\epsilon}{1-3\epsilon} +V_{ijk}\,V_j\,V_{kl}V_l\,\frac{3\,(1-7\epsilon)}{1-3\epsilon} +\\ \nonumber && V_{ijkl}\,V_j\,V_k\,V_l\,\frac{(1-5\epsilon)\,(1-7\epsilon)}{(1-3\epsilon)^2}\,\Big)\, F_1\,F_{4-d_T}\,F_{7-2d_T}\,F_{10-3d_T}\,|\vec{p}^{\, T}|^{2(4d_T-12)}\,,
\end{eqnarray}
 where the coefficients are defined in appendix \ref{sect:integrals}.
 
\subsection{Renormalization and $\beta $ functions}\label{subsec:renormqft}

For the sake of clarity, in the following we will first describe the
case of $d=4-\epsilon$ dimensions in detail. Expanding the $\tilde{\phi}_i$ in $d_T=3-\epsilon$, one finds

\begin{equation}
\tilde{\phi}_i=C_0\,(1+C_1\,\log|\vec{p}^{\, T}|+C_2\,(\log|\vec{p}^{\, T}|)^2+\cdots)\,.
\end{equation}
The $C_i$'s are divergent as $\epsilon\rightarrow 0$. These divergences can be renormalized introducing a renormalized coupling $u_i$ by demanding that $C_0$ is finite. Restoring the powers of the scale, one finds
\begin{eqnarray}
 \nu_i= && \mu^{\frac{\epsilon}{2}}\,\Big(u_i+\alpha^{(1)}V_i+\alpha^{(2)}\,V_{ij}\,V_j+\alpha^{(3)}_1\,V_{ijk}V_jV_k+\alpha^{(3)}_2\,V_{ij}V_{jk}V_k\\ \nonumber && +\alpha^{(4)}_1\,V_{ij}\,V_{jk}V_{kl}\,V_l+\alpha^{(4)}_2\,V_{ij}\,V_{jkl}\,V_k\,V_l +\alpha^{(4)}_3\,V_{ijk}\,V_j\,V_{kl}V_l+\alpha^{(4)}_4\,  V_{ijkl}\,V_j\,V_k\,V_l\Big)\,,
\end{eqnarray}
where the RHS is evaluated at $u_i$, and 

\begin{equation}
\begin{array}{l c l}
\alpha^{(1)}=\frac{\Omega}{\epsilon} & & \alpha^{(2)}=\frac{\Omega^2}{2\epsilon^2}-\frac{\Omega^2}{\epsilon} \\
\alpha^{(3)}_1=\frac{\Omega^3}{6\epsilon^3}-\frac{\Omega^3}{3\epsilon^2}-\frac{\Omega^3}{3\epsilon} & & \alpha^{(3)}_2=\frac{\Omega^3}{6\epsilon^3}-\frac{\Omega^3}{\epsilon^2}-\frac{8\Omega^3}{3\epsilon} \\
\alpha^{(4)}_1=\frac{\Omega^4}{24\epsilon^4}-\frac{\Omega^4}{2\epsilon^3}+\frac{19\Omega^4}{6\epsilon^2}-\frac{10\Omega^4}{\epsilon} & & \alpha^{(4)}_2=\frac{\Omega^4}{24\epsilon^4}-\frac{\Omega^4}{4\epsilon^3}+\frac{5\Omega^4}{12\epsilon^2}+\frac{\Omega^4}{12\epsilon}  \\ 
\alpha^{(4)}_3=\frac{\Omega^4}{8\epsilon^4}-\frac{2\Omega^4}{3\epsilon^3}+\frac{\Omega^4}{\epsilon^2}+\frac{11\Omega^4}{6\epsilon}  && \alpha^{(4)}_4=\frac{\Omega^4}{24\epsilon^4}-\frac{\Omega^4}{12\epsilon^3}-\frac{\Omega^4}{12\epsilon^2}-\frac{\Omega^4}{12\epsilon} 
\end{array}
\end{equation}
Here, as in \cite{Rodriguez-Gomez:2022gif}, $\Omega=\frac{1}{32\pi^2}$.

Demanding that the bare coupling is independent on the renormalization scale (and using that the $\beta $ function for the bulk couplings is $\beta_{g_{\alpha}}=-\epsilon\,g_{\alpha}$) we find the $\beta $ functions for the defect couplings:
\begin{eqnarray}
\label{eq:betainQFT}
 \beta_i= && -\frac{\epsilon}{2}\,u_i+\beta^{(1)}V_i+\beta^{(2)}\,V_{ij}\,V_j+\beta^{(3)}_1\,V_{ijk}V_jV_k+\beta^{(3)}_2\,V_{ij}V_{jk}V_k\\ \nonumber && +\beta^{(4)}_1\,V_{ij}\,V_{jk}V_{kl}\,V_l+\beta^{(4)}_2\,V_{ij}\,V_{jkl}\,V_k\,V_l +\beta^{(4)}_3\,V_{ijk}\,V_j\,V_{kl}V_l+\beta^{(4)}_4\,  V_{ijkl}\,V_j\,V_k\,V_l\,,
\end{eqnarray}
with
\begin{eqnarray}
&& \beta^{(1)}=2\Omega  \ ,\qquad \beta^{(2)}=-4\Omega^2 \ ,\qquad
\beta^{(3)}_1=-2\Omega^3 \ ,\ \ \  \beta^{(3)}_2=16\Omega^3\ ,
\nonumber\\
&& \beta^{(4)}_1=-80\Omega^4 \, \quad \beta^{(4)}_2=\frac{2\Omega^4}{3}\ ,\ \qquad 
\beta^{(4)}_3=\frac{44\Omega^4}{3} \ ,\quad  \beta^{(4)}_4=-\frac{2\Omega^4}{3}\ .
\label{coefi}
\end{eqnarray}

A similar calculation in $d=6-\epsilon $ dimensions
gives $\beta$ functions with the same structure as in 
\eqref{eq:betainQFT}. To three-loop order, the coefficients are now given by

\begin{equation}
\label{betaseis}
 \beta^{(1)}=\Omega  \ ,\qquad \beta^{(2)}=-\frac12 \Omega^2 \ ,\qquad
\beta^{(3)}_1=-\frac18 \Omega^3 \ ,\ \ \  \beta^{(3)}_2=\frac12 \Omega^3\ ,
\end{equation}
where, for the 6d theory, we define $\Omega =\frac{1}{8\pi^2}$.

The  coefficients $\beta^{(1)}$ and $\beta^{(2)}$ in \eqref{coefi} and \eqref{betaseis} reproduce the two-loop terms previously computed in  \cite{Rodriguez-Gomez:2022gif}, and also agree with the earlier calculations in \cite{Allais:2014fqa,Cuomo:2021kfm}  once the double-scaling limit is taken ({\it c.f.}
eq. (19) in \cite{Allais:2014fqa}  and eq. (3.17) in \cite{Cuomo:2021kfm}).

\subsection{Fixed points and dimensional disentanglement}

Let us  consider the four-loop $\beta $ functions \eqref{eq:betainQFT}. One can check that all solutions of $\beta_i=0$ are of the form

\begin{equation}
\label{ansf}
u_i^{\star}=F_i^{\rm one-loop}(g_{\alpha})\, \sqrt{f(\epsilon)}\,,
\end{equation}
that is, the $\epsilon$-dependence factorizes, 
with 
\begin{equation}
\label{f}
f(\epsilon)=\epsilon+\frac{3}{2}\epsilon^2+\frac{3}{8}\epsilon^3+\frac{11}{16}\epsilon^4+\cdots\,.
\end{equation}
and
$F_i^{\rm one-loop}(g_{\alpha})$ completely determined by the one-loop $\beta_i=0$ equation,
$\epsilon u_i=4\Omega\,V_i$.
This suggests that the location of the fixed points of the defect theory {\it to all orders} is determined by the vanishing of the one-loop  $\beta $ function up to a universal overall function $f$ which entirely encodes the $\epsilon$ dependence (and hence the dimension), a phenomenon which was dubbed \textit{dimensional disentanglement} (DD) in \cite{Rodriguez-Gomez:2022gif}. 
The $F_i^{\rm one-loop}(g_{\alpha})$ is given in terms of ratios of bulk couplings. These ratios are RG invariants, since all couplings $g_\alpha $ have the same classical flow, $\beta_{g_{\alpha}}=-\epsilon g_{\alpha }$.

The four-loop check extends the conjecture of \cite{Rodriguez-Gomez:2022gif} to the general class \eqref{trea} of scalar field models with defects. Surprisingly,  we find that the function $f(\epsilon)$ is universal: it is the same function for any 4d scalar field theory of the type \eqref{trea}. Although we have focused on the four-dimensional models, DD  in the fixed points also occurs in  6d scalar field models with defect of the form \eqref{trea}, a property which, as shown below, only holds in the double-scaling limit. In the 6d case, the function $f(\epsilon)$ is different, and one finds an expansion of the form $f_{6d}(\epsilon)=\epsilon^2\,(1+\epsilon+\cdots)$.

\medskip

To understand the origin of DD, it is useful 
to derive the solutions of $\beta_i=0$ in detail.
Let us write
 \begin{equation}
 u_i=\epsilon^{\frac{1}{2}} \left(a_i\,+b_i\,\epsilon +c_i\,\epsilon^{2}+\ldots\right)
 \end{equation}
Then
   \begin{eqnarray}
 && \beta_i=\Big(-\frac{\epsilon^{\frac{3}{2}}}{2}\,a_i+\beta^{(1)}V_i\Big)+\Big(-\frac{\epsilon^{\frac{5}{2}}}{2}\,b_i+\beta^{(1)}V_{ij}\,b_j\,\epsilon^{\frac{3}{2}}+\beta^{(2)}\,V_{ij}\,V_j\Big)+ \nonumber \\ \nonumber &&
 +\Big[-\frac{\epsilon^{\frac{7}{2}}}{2}c_i+\beta^{(1)}V_{ij}\,c_j\,\epsilon^{\frac{5}{2}}+\frac{\beta^{(1)}}{2}\,V_{ijk}\,b_j\,b_k\,\epsilon^3 
\nonumber \\ \nonumber && +\big(\beta^{(2)}\,V_{ijk}\,V_j+\beta^{(2)}\,V_{ij}\,V_{jk})\,b_k\epsilon^{\frac{3}{2}}+\beta^{(3)}_1\,V_{ijk}V_jV_k+\beta^{(3)}_2\,V_{ij}V_{jk}V_k
\Big]\,.
 \end{eqnarray}
 Here everything is assumed to be evaluated at $a_i\,\epsilon^{\frac{1}{2}}$. The vanishing of the $\beta $ functions implies a cancellation order by order. A crucial property in the derivation is the homogeneity of $V$, which implies the general relation
 \begin{equation}
     u_j V_{i_1 ...,i_p j}=(n-p) V_{i_1 ...i_p}\ .
 \end{equation}
 From the leading term, we find
 \begin{equation}
 \frac{\epsilon^{\frac{3}{2}}}{2}\,a_i=\beta^{(1)}V_i\,.
 \end{equation}
Using this and the homogeneity of $V$, the second term gives
 the relation
  \begin{equation}
 0= -\frac{\epsilon^{\frac{5}{2}}}{2}\,b_i+\beta^{(1)}\,V_{ij}\,b_j\,\epsilon^{\frac{3}{2}}+\frac{3\beta^{(2)}}{2\beta^{(1)}}\,V_{i}\,\epsilon\,.
 \end{equation}
This can be easily solved by choosing $b_i=m\,a_i$ for some $m$, since, by virtue  of the homogeneity of $V$,
   \begin{equation}
 0= -\frac{\epsilon^{\frac{5}{2}}}{2}\,m\,a_i+3\,(m\,\beta^{(1)}+\frac{\beta^{(2)}}{2\beta^{(1)}})\,V_{i}\,\epsilon\,.
 \end{equation}
 Using again the leading order equation, we obtain
 
   \begin{equation}
 0= -\frac{m}{2}+\frac{3}{2\beta^{(1)}}\,(m\,\beta^{(1)}+\frac{\beta^{(2)}}{2\beta^{(1)}})\,.
 \end{equation}
This yields
$m= \frac{3}{4}$,
in agreement with the expansion of \eqref{f}.
As for the last term, let us also assume $c_i=\kappa\,a_i$. Then
 \begin{eqnarray}
&& -\frac{\epsilon^{\frac{7}{2}}}{2}\kappa\ a_i+\beta^{(1)}V_{ij}\,\kappa a_j\,\epsilon^{\frac{5}{2}}+\frac{\beta^{(1)}}{2}\,V_{ijk}\,m^2 a_j\,a_k\,\epsilon^3 
\nonumber\\
&&+\big(\beta^{(2)}\,V_{ijk}\,V_j+\beta^{(2)}\,V_{ij}\,V_{jk})\,m a_k\epsilon^{\frac{3}{2}}+\beta^{(3)}_1\,V_{ijk}V_jV_k+\beta^{(3)}_2\,V_{ij}V_{jk}V_k=0\,.
 \end{eqnarray}
 Using the homogeneity of $V$
  \begin{equation}
 -\frac{\epsilon^{\frac{7}{2}}}{2}\kappa \ a_i+3\,\beta^{(1)}V_{i}\,(\kappa +m^2) \epsilon^2+5\,\beta^{(2)}\,V_{ij}\,V_j\,m\,\epsilon+\beta^{(3)}_1\,V_{ijk}V_jV_k+\beta^{(3)}_2\,V_{ij}V_{jk}V_k=0\,.
 \end{equation}
Using the leading order equation and the homogeneity of $V$ we find 
 \begin{equation}
 -\frac{\kappa}{2}+ \frac{3}{2}\,(\kappa +m^2) +  \frac{15\,\beta^{(2)}}{4(\beta^{(1)})^2}\,m+ \frac{6\beta^{(3)}_1+9\,\beta^{(3)}_2}{8(\beta^{(1)})^3}=0\,.
 \end{equation}
 Solving for $\kappa $ we get
$ \kappa =-\frac{3}{32}$, 
 once again in  agreement with the expansion of \eqref{f}.

\smallskip

\begin{figure}[h!]
\centering
\includegraphics[scale=.3]{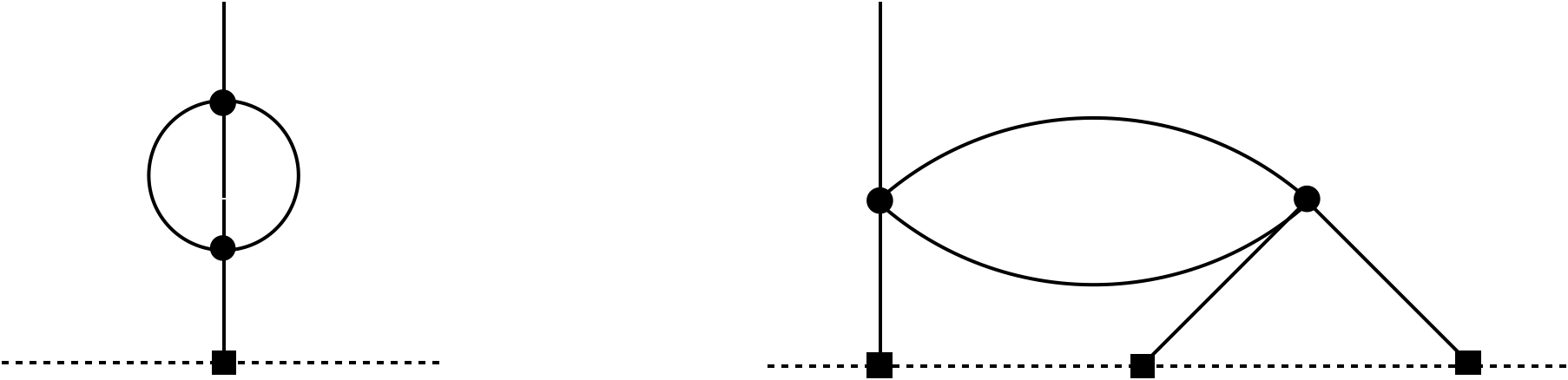}
\caption{Leading bulk loop diagrams contributing to the $\beta$ functions in 4d.   Black circles are ${g}$ vertices while squares represent the $\nu$ coupling to the defect.}
\label{fig:loops}
\end{figure}

Even though we have so far explicitly shown dimensional disentanglement up to four loops in general theories, it is clear that the strategy extends to arbitrary orders. To further understand DD it is enlightening to study when it fails to hold, as happens upon including  bulk loops. Focusing for definiteness on $d=4-\epsilon$, where $V$ is quartic, these enter to order $\mathcal{O}(V^2)$, with the diagrams in figure \ref{fig:loops}. 

The first diagram contributes to the anomalous dimension of $\phi$. The second diagram would produce a term in the $\beta$ function of the form

\begin{equation}
\delta \beta_i=\delta\beta^{(2)}\, V_{ijk}V_{jk}\,.
\end{equation}
 Thus, to this order the full $\beta $ functions would be

\begin{equation}
 \beta_i=  -\frac{\epsilon}{2}\,u_i+\beta^{(1)}V_i+\beta^{(2)}\,V_{ij}\,V_j+\delta\beta^{(2)}\, V_{ijk}V_{jk}\,.
\end{equation}
We already see a crucial difference: while in the large charge limit, the $\beta $ function to order $\mathcal{O}(V^k)$ contains a total of $2k-1$ derivatives, the corrections 
(coming from bulk loops) contain, to order $\mathcal{O}(V^k)$, more derivatives. For instance, the leading correction to order $\mathcal{O}(V^k)$ contains $2k+1$ derivatives. To see the implications of this, let us proceed as before and assume
 \begin{equation}
 u_i= \epsilon^{\frac{1}{2}}\left( a_i +b_i\,\epsilon+...\right)\,.
 \end{equation}
Then
\begin{equation}
 \beta_i=  -\frac{\epsilon^{\frac{3}{2}}}{2}\,a_i -\frac{\epsilon^{\frac{5}{2}}}{2}\,b_i+\beta^{(1)}V_i+\beta^{(1)}V_{ij}\,b_j\,\epsilon^{\frac{3}{2}}+\beta^{(2)}\,V_{ij}\,V_j+\delta\beta^{(2)}\, V_{ijk}V_{jk}\,,
\end{equation}
where again everything is evaluated at $a_i\epsilon^{\frac{1}{2}}$. Grouping terms with the same dependence of $\epsilon$ we see that
\begin{equation}
 \beta_i=  \Big(-\frac{\epsilon^{\frac{3}{2}}}{2}\,a_i+\beta^{(1)}V_i+\delta\beta^{(2)}\, V_{ijk}V_{jk}\Big) +\Big(-\frac{\epsilon^{\frac{5}{2}}}{2}\,b_i+\beta^{(1)}V_{ij}\,b_j\,\epsilon^{\frac{3}{2}}+\beta^{(2)}\,V_{ij}\,V_j\Big)\,.
\end{equation}
We now find the leading equation
\begin{equation}
\frac{\epsilon^{\frac{3}{2}}}{2}\,a_i=\beta^{(1)}V_i+\delta\beta^{(2)}\, V_{ijk}V_{jk}\,.
\end{equation}
The crucial difference is that now there is an extra term with higher powers of the bulk coupling constant. We can solve  this equation in perturbation theory, finding

\begin{equation}
a_i=a_i^{\circ}+2\,\epsilon^{-\frac{3}{2}}\,\delta\beta^{(2)}\,V^{\circ}_{ijk}V^{\circ}_{jk}\,,
\end{equation}
where $a_i^0$ is the solution to $\frac{\epsilon^{\frac{3}{2}}}{2}\,a_i=\beta^{(1)}V_i$ and $V^{\circ}_{ijk}=V_{ijk}(u_i^\circ)$,
with
$u_i^\circ  =a_i\epsilon^{\frac{1}{2}}$.

Thus we see that upon including loop corrections, the fixed points will be of the form

\begin{equation}
u_i^*=a_i^{\circ}\,\sqrt{\epsilon} \sum_{k=0}^\infty f_{i,k}\,\epsilon^k\,,
\end{equation}
where the $\{ f_{i,k}\} $ are non-trivial functions of the bulk couplings.
They are of the form
\begin{equation}
f_{i,k}= c_k F_i(g_\alpha)\big[1+\mathcal{O}(g_{\alpha}^{k+1})\big]\,.
\end{equation}
In the double-scaling limit, $1+\mathcal{O}(g_{\alpha}^{k+1})\to 1$  and
the function $F_i(g_\alpha)$ factorizes, with $F_i(g_\alpha)=F_i^{\rm one-loop}(g_{\alpha})$, giving rise  to dimensional disentanglement.

\smallskip

Summarizing, dimensional disentanglement is tied to the fact that, in the double scaling limit, to any given order in the bulk couplings only terms with the same number of derivatives of the potential with respect to the fields appear.
This is no longer true in the full quantum theory once bulk loops are included.
DD arises also thanks to the homogeneity of the potential (it is a degree $n$ polynomial in the fields, linear in bulk couplings, with $n=4$ in 4d and $n=3$ in 6d).

In   \cite{Rodriguez-Gomez:2022gif} it was shown that, up to two-loop order, the   $\beta $ functions can be obtained as a gradient
from a function ${\cal H}$, that is,
$\beta_i=2\partial_i {\cal H}$. 
Although the four-loop $\beta $ functions in \eqref{eq:betainQFT} are not the gradient of any function, nevertheless dimensional disentanglement still holds, due to the structure of the corrections described above.
Beyond two loops, the $\beta $-functions have a scheme-dependence and, as discussed below in section \ref{esquema}, it is possible to choose a scheme where they are still given as a gradient function.

\subsection{Some physical implications of DD}

DD implies that fixed points have the form \eqref{ansf}.
The main physical consequence is that the positions of fixed points in the defect coupling space do not depend
on $\epsilon $ modulo an overall scale given by $f(\epsilon)$.
In other words their relative position is independent of
the dimension.

The RG flow, however, can have a dependence on the dimension, despite the fact that fixed points do not move when $\epsilon $ is varied, except for an overall scale.
The way this happens can be illustrated by the twins model
discussed in \cite{Rodriguez-Gomez:2022gif}.
It is defined by the action ($d=4-\epsilon$)
\begin{equation}
\label{gemelos}
\mathcal{S}= \int d^d x\left( \frac{1}{2}\big(\partial\phi_1\big)^2+\frac{1}{2}\big(\partial\phi_2\big)^2+V(\phi_1,\phi_2)-\nu_i\,\phi_i\,\delta_T(\vec{x})\right)\,, \quad i=1,2\ .\, , 
\end{equation}
with
\begin{equation}
    V(\phi_1,\phi_2)=\frac14 g_1\phi_1^4+\frac14 g_2\phi_2^4+\frac12 g_3 \phi_1^2\phi_2^2\ .
\end{equation}
The  $\beta $
functions for the defect couplings can be read from \eqref{eq:betainQFT}.
To quadratic order in the couplings, they are given by
\cite{Rodriguez-Gomez:2022gif}
\begin{eqnarray}
\beta_{\nu_1}&=& u_{1}\Big( -\frac{\epsilon}{2}+2\, \,\Omega\,(g_1\, u_{1}^2+g_3\,u_{2}^2)
\nonumber\\
&-&
4\,\Omega^2\, \big[ 3g_1^2\,u_{1}^4 +2g_3(2g_1+g_3)\,u_{1}^2\,u_{2}^2+g_3(2g_2+g_3) \,u_{2}^4\big]\Big)\, ,
\end{eqnarray}
\begin{eqnarray}
\beta_{\nu_2}&=& u_{2}\Big( -\frac{\epsilon}{2}+2\, \,\Omega\,(g_2\,u_{2}^2+g_3\,u_{1}^2)
\nonumber\\
&-&
4\,\Omega^2\, \big[ 3g_2^2\,u_{2}^4 +2g_3(2g_2+g_3)\,u_{1}^2\,u_{2}^2+g_3(2g_1+g_3) \,u_{1}^4\big]\Big)\,.
\label{betagem}
\end{eqnarray}
Defining
\begin{equation}
 x_1=\Omega\, g_1 \,u_{1}^2\ ,\quad   x_2=\Omega\, g_2 \,u_{2}^2\, ;\qquad \zeta=\frac{g_3}{g_1}\ ,\ \ 
  \eta=\frac{g_3}{g_2}\, ;
\end{equation}
one finds that fixed points $(x_1^*,x_2^*)$ are located at

\begin{equation}
\label{fijoabb}
 (x_1^*,x_2^*):\ \qquad    a)\ \,(0,\,0), \qquad b)\ \,(0,\,\frac{1}{4} ) f(\epsilon)\qquad b')\,, \,(\frac{1}{4},\,0)f(\epsilon)\, ,
\end{equation}
\begin{equation}
\label{fijoc}
    c)\ \, \left(\frac{1-\eta}{4(1-\zeta\eta)}, \, \frac{1-\zeta}{4(1-\zeta\eta)} \right)f(\epsilon)\,;\qquad f(\epsilon)=\epsilon+\frac{3}{2}\epsilon^2+\cdots\,.
\end{equation}
In \cite{Rodriguez-Gomez:2022gif}, IR stability was studied only to linear order in $\epsilon $. To understand to what extent quantitative and qualitative features of the RG flow can depend on $\epsilon$, it is important to extend the stability analysis to order $\epsilon^2$. Consider the RG time variable $t=-\log\mu $. Perturbing around the fixed points we find the following eigenvalues $(\lambda_1,\lambda_2)$ of the Hessian: 
\begin{equation}
 (\lambda_1,\lambda_2):\qquad   a)\, \ \left(\frac{\epsilon}{2},\,\frac{\epsilon}{2}\right),\qquad b)\,\ \frac12 \left(-\epsilon+\frac32 \epsilon^2,\,\epsilon (1-\eta)\big(1-\frac12\epsilon\, \eta \big)\right),
 \end{equation}
\begin{equation}
b')\,\  \frac12 \left(-\epsilon+\frac32 \epsilon^2,\, \epsilon(1-\zeta)\big( 1-\frac12 \epsilon\, \zeta\big)\right)\,,
\end{equation}
\begin{equation}
    c)\, \ \frac12 \left(-\epsilon\,\frac{(1-\eta)\,(1-\zeta)}{1-\eta\,\zeta}\left(1-\frac{\epsilon}{2}\,  \frac{3-2\zeta-2\eta+\zeta\eta}{1-\zeta\eta}\right),\,-\epsilon+\frac32 \epsilon^2\right)\,.
\end{equation}
IR stability of a given fixed point requires that both eigenvalues are negative. 

We see that the $\epsilon$ dependence does not factorize.
Stability properties  change by varying $\epsilon$ at fixed
couplings $(\zeta,\eta)$. For example, taking $\eta\gg 1$, the $b)$ fixed point is stable  for sufficiently small $\epsilon$, but it becomes unstable when $\epsilon>2/\eta$. This implies a drastic change in the RG flow, despite the fact that the relative positions of fixed points remain unchanged: an attractive fixed point becomes repulsive as $\epsilon$ is increased above a critical value (while keeping $\epsilon\ll 1$).

\medskip

In conclusion, in the double-scaling limit, on one hand, fixed points satisfy the DD property, which allows one to determine them exactly (modulo the overall numerical constant $f(\epsilon)$) by a one-loop calculation. On the other hand, $\beta$ functions still describe extremely rich RG flows  exhibiting phenomena such as fixed point creation/annihilation and non-trivial dynamics as $\epsilon$ is varied.

\section{Alternative calculation of $\beta$-functions}\label{sec:betas}

In this section --in which we will set $\epsilon=0$, that is, we shall compute the $\beta $-functions of the defect couplings for  $d=4,6$ (this means that in our convention $d_T=3,4$ respectively)-- we will show that the $\beta $ functions can be computed in an elegant way from corrections to the Coulomb potential. In the appendix \ref{app:Rigidholography} a similar calculation of the 
$\beta $ functions will be given using rigid holography. 

\subsection{$\beta$ function for the defect couplings}

Let us start with the  action \eqref{pres}. Recall that $V$ is a homogeneous polynomial of  the fields $\phi_i$ of degree $n=4$ in the $d=4$ theory, and $n=3 $ in the $d=6$ theory.

We shall use spherical coordinates, and place the defect at $r=0$. Explicitly

\begin{equation}
ds^2=d\vec{x}_{||}^2+d\vec{x}_T^2=d\vec{x}_{||}^2+dr^2+r^2\,d\Omega_{d_T-1}^2\,.
\end{equation}

For our purposes, it is sufficient to consider spherical symmetric solutions, where  $\phi_i$ only depends on $r$. Under this assumption, the equation of motion reads

\begin{equation}
\label{equV}
\partial_r\big({r^{d_T-1}}\partial_r\phi_i\big)-r^{d_T-1}\,V_i=-\nu_i \delta_T\,,
\end{equation}
where $V_i=\frac{\partial V}{\partial\phi_i}$. Writing

\begin{equation}
\phi_i=\frac{u_i}{r^{d_T-2}},
\end{equation}
the equation of motion away from the source becomes

\begin{equation}
\label{eq:eomradial}
\partial_r\big(r^{3-d_T}\,\partial_r u_i\big)-\frac{1}{r^{d_T-1}}\,V_i=0\,,
\end{equation}
where $V_i$ is now evaluated at $u_i$. We can solve this equation in perturbation theory by setting
\begin{equation}
    u_i=s_i+f^{(1)}_i(r)+f^{(2)}_i(r)+\cdots\,,
\end{equation}
where $f^{(k)}$ is of order $g_\alpha^k$  and $s_i$ is a constant. Up to order 3
\begin{eqnarray}
&& \partial_r\big(r^{3-d_T}\partial_r f_i^{(1)}\big)+\partial_r\big(r^{3-d_T}\partial_r f_i^{(2)}\big)+\partial_r\big(r^{3-d_T}\partial_r f_i^{(3)}\big)
\nonumber\\ \nonumber\\
&&-\frac{1}{r^{d_T-1}}\,\Big\{ V_i+V_{ij}\,f^{(1)}_j+\big(V_{ij}f_j^{(2)}+\frac{1}{2}V_{ijk}f_j^{(1)}f^{(1)}_k\big)\Big\}=0\,,
\nonumber
\end{eqnarray}
where  $V$ and its derivatives are now evaluated at $s_i$. It is  straightforward to solve this equation order by order,
finding
\begin{equation}
\label{eq:fpert}
\begin{aligned}
f_i^{(1)}&=-\frac{V_i}{d_T-2}\,\log r\,,\qquad f_i^{(2)}=\frac{V_{ij}V_j}{2\,(d_T-2)^3}\,\big(2\log r+(d_T-2)\,(\log r)^2\big)\,,\\
f_i^{(3)}&=- \frac{V_{ij} V_{jk} V_k}{6 (d_T-2)^5}\left (  (d_T-2)^2 (\log r)^3+6  (d_T-2) (\log r)^2 +12 \log r\right)-\\
&- \frac{V_{ijk} V_j V_k}{6 (d_T-2)^5} \left(  (d_T-2)^2 (\log r)^3+3 (d_T-2) (\log r)^2+6 \log r \right)\,.
\end{aligned}
\end{equation}

The constants $s_i$ can be determined from the $\delta_T$ source term on the right hand side of the equations of motion 
\eqref{equV}. They are given by
\begin{equation}
\label{normas}
s _i^{d=4}=\frac{\nu_i}{4\pi}\,, \quad\quad 
s _i^{d=6}=\frac{\nu_i}{4\pi^2}\ .
\end{equation}

The charges $u_i$ can be viewed as a  ``running" version of $\nu_i$. To third order
\begin{equation}
\begin{aligned}
u_i(r)&=\nu_i-\frac{2\Omega\, V_i}{d_T-2}\,\log r+\frac{2\Omega^2 V_{ij}V_j}{(d_T-2)^3}\,\big(2\log r+(d_T-2)\,(\log r)^2\big) - \\
&- \frac{4\Omega^3 V_{ij} V_{jk} V_k}{3 (d_T-2)^5 }\left (  (d_T-2)^2 (\log r)^3+6  (d_T-2) (\log r)^2 +12 \log r\right)-\\
&- \frac{4 \Omega^3 V_{ijk} V_j V_k}{3 (d_T-2)^5 } \left(  (d_T-2)^2 (\log r)^3+3 (d_T-2) (\log r)^2+6 \log r \right)\,,
\end{aligned}
\end{equation}
where now it is understood that $V$ and its derivatives are evaluated at $s_i$. The numerical constant $\Omega$ was introduced in section \ref{subsec:renormqft}
($\Omega=\frac{1}{32\pi^2}$ in $d=4$; $\Omega=\frac{1}{8\pi^2}$ in $d=6$).

Inverting this formula, we get

\begin{equation}
\begin{aligned}
\nu_i=&u_i+\frac{2\Omega\, V_i}{d_T-2}\,\log r-\frac{2\Omega^2 V_{ij}V_j}{(d_T-2)^3}\,\big(2\log r-(d_T-2)\,(\log r)^2\big) +\\
&+\frac{4\Omega^3 V_{ij} V_{jk} V_k}{3 (d_T-2)^5}\left((d_T-2)^2 (\log r)^3-6(d_T-2) (\log r)^2+12 \log r\right)-\\
&+\frac{4\Omega^3V_{ijk} V_{j} V_k}{3 (d_T-2)^5} \left( (d_T-2)^2 (\log r)^3-3 (d_T-2) (\log r)^2+6 \log r \right)\,.
\end{aligned}
\end{equation}
with $V$ and its derivatives being evaluated at $s_i$. Interpreting $r^{-1}$ as the RG scale, we can compute the $\beta $ function for $u_i$
\begin{equation}
    \beta_i =- \frac{\partial u_i}{\partial \log r}\ ,
\end{equation}
by imposing the scale-independence ($r$-independence) of the ``bare coupling" $\nu_i$. We obtain

\begin{equation}
\label{eq:betainholography}
\beta_i=2\,c\,\Omega\,V_i-4\,c^3\,\Omega^2\,V_{ij}V_j+8\,c^5\,\Omega^3\,(V_{ijk}V_j\,V_k+2\,V_{ij}V_{jk}\,V_k)\, , 
\end{equation}
with $c=1/(d_T-2)$ (hence $c=1$ in $d=4$ and $c=1/2$ in $d=6$).
Up to second order, this formula exactly matches the quantum field theory results given in \eqref{eq:betainQFT}, \eqref{coefi}, \eqref{betaseis}, for the 4d and 6d theories. As we will shortly review, this is to be expected, since only the one-loop and two-loop terms of the $\beta$ function are expected to be scheme-independent.

 Using this same method it is straightforward --albeit tedious-- to go to higher loops. The four loop contribution is  derived in appendix \ref{sect:holo4}. Remarkably, the $\beta$ function is a gradient flow in the defect coupling space, 
 \begin{equation}
 \label{eq:hgrad}
     \beta_i=2\,c\,\partial_i\mathcal{H}\,,
 \end{equation}
 where
 \begin{equation}
     \mathcal{H}=\Omega\,V-c^2\,\Omega^2\,V_i^2+4\,c^4\,\Omega^3\,V_{jk}V_jV_k-8 c^6\,\Omega^4\, V_{ijk} V_i V_{j} V_k-20 c^6\,\Omega^4\, V_i V_{ij} V_{jk} V_k\,.
 \end{equation}
This supports the conjecture made in \cite{Rodriguez-Gomez:2022gif}, albeit in a particular scheme which coincides with the one implicitly chosen by this alternative method.

\label{esquema}
\subsection{Changing scheme}

From 3-loops on, the coefficients in the $\beta $ function obtained through the previous method fail to match the corresponding coefficients in the field-theoretic result in \eqref{eq:betainQFT}. For example, in the four-dimensional theory,  in \eqref{coefi} and \eqref{betaseis}, $\beta^{(3)}_1=-2 c^4\Omega^3$ whereas in \eqref{eq:betainholography}, $\beta^{(3)}_1=16 c^5\Omega^3$. We note that the coefficient $\beta^{(3)}_2$ is the same in both calculations.

It is well known that $\beta $ functions are scheme-dependent beyond two loops. To understand the origin of the discrepancy in more detail, let us study the effect of changing the scheme.  To do this,  we redefine our $u_i$ couplings in terms of  new couplings $\tilde{u}_i$. A natural ansatz is
\begin{equation}
u_i=\tilde{u}_i+\alpha_1\,\tilde{V}_i+\alpha_2\,\tilde{V}_{ij}\,\tilde{V}_j\,,
\end{equation}
where $\tilde{V}$ means $V$ evaluated on the $\tilde{u}_i^R$'s. Then
\begin{equation}
\beta_{u_i}=\Big(\delta_{ji}+\alpha_1\,\tilde{V}_{ij}+\alpha_2\,(\tilde{V}_{ijk}\tilde{V}_k+\tilde{V}_{ik}\tilde{V}_{kj})\Big)\beta_{\tilde{u}_j}\, .
\end{equation}
Inverting this matrix. we get
\begin{equation}
\beta_{\tilde{u}_i}=\Big(\delta_{il}-\alpha_1\,\tilde{V}_{il}-(\alpha_2\, \tilde{V}_{ilk}\tilde{V}_k+(\alpha_2-\alpha_1^2)\,\tilde{V}_{ik}\tilde{V}_{kl})\Big)\,\beta_{u_l}\ .
\end{equation}
In turn
\begin{eqnarray}
&&\beta_{u_i}=2\,c\,\Omega\,\tilde{V}_i+(2\,c\,\alpha_1\,\Omega-4\,c^3\,\Omega^2)\,\tilde{V}_{ij}\tilde{V}_j\\ \nonumber && +(2\,c\,\alpha_2\,\Omega-4\,c^3\,\alpha_1\Omega^2+16\,c^5\,\Omega^3)\,\tilde{V}_{ij}\tilde{V}_{jk}\,\tilde{V}_k+(8\,c^5\,\Omega^3-4\,c^3\,\alpha_1\Omega^2)\,\tilde{V}_{ijk}\tilde{V}_j\tilde{V}_k\,,
\end{eqnarray}
Therefore

\begin{equation}
\beta_{\tilde{u}_i}=2\,c\,\Omega\,\tilde{V}_i-4\,c^3\,\Omega^2\,\tilde{V}_{ij}\tilde{V}_j+\beta_1^{(3)}\,\tilde{V}_{ijk}\tilde{V}_j\tilde{V}_k +\beta_2^{(3)}\,\tilde{V}_{ij}\tilde{V}_{jk}\,\tilde{V}_k\,,
\end{equation}
with

\begin{equation}
\beta_1^{(3)}=8c^5\Omega^3-4c^4\Omega^2\alpha_1-2c\Omega\alpha_2\,,\qquad \beta_2^{(3)}=16c^5\Omega^3\,.
\end{equation}
Thus we see that  the coefficient $\beta^{(3)}_1$ where we find a disagreement between \eqref{eq:betainQFT} and \eqref{eq:betainholography} is precisely  that  altered by  redefinition of couplings.

\section{Instabilities in  defect field theories}\label{sec:Instabilities}

In the previous sections we have computed the $\beta$ functions for the defect couplings assuming a double-scaling limit. In particular, the effect of such limit is to freeze the running of the bulk couplings, in such a way that the bulk theory is effectively a CFT if we set $\epsilon=0$ (so that the classical running is also frozen). Thus, armed with our previous results, we will now study cases where also the defect 
$\beta $ functions vanish, so that we have a defect CFT (dCFT). 
It is of interest to investigate if these potential dCFT's may have further instabilities triggered by condensates of marginal or relevant operators, just as it happens in
 the scalar QED example of \cite{Aharony:2022ntz} (see also appendix \ref{subsec:holoscalarqed}), where the dCFT ceases to exist beyond certain critical values of the couplings.

To make this concrete, let us consider a model with two fields $\rho$ and $\vec{\phi}$, being $\vec{\phi}$ an $O(N)$ vector. We choose the potential to be of the form $\rho^{n-2}\,\vec{\phi}^2$, with $n=4$ in $d=4$ and $n=3$ in $d=6$. Introducing now a defect to which in general both $\rho$ and $\phi^i$ couple, the $\beta$-functions of the defect couplings can be computed from
\eqref{eq:betainQFT}. Denoting the corresponding renormalized defect couplings by $u_{\rho}$ and $u_{\phi^i}$ in the obvious way, it is straightforward to check  that the model has a fixed point at $u_{\phi^i}=0$ for arbitrary $u_{\rho}$.\footnote{Note that the $d=4$ model is a particular case of the twins model \eqref{gemelos} with $g_1=g_2=0$, with the $\beta$'s given in \eqref{betagem}.} The action is given by (we denote the only bare defect coupling simply by $\nu$)

\begin{equation}
S_{\rm eff}=\int d^dx\, \left(\frac{1}{2}(\partial\vec{\phi})^2+\frac{1}{2}(\partial\rho)^2+g\,\rho^{n-2}\,\vec{\phi}^2-\nu\,\rho\,\delta_T \right)\,,
\end{equation}
which describes, in principle, a dCFT. We wish to study whether, similarly to the QED case in \cite{Aharony:2022ntz}, there are other instabilities triggered by  relevant operators.

\subsection{One-loop considerations in field theory}

In this subsection we shall  analyze the stability of the fixed points
in perturbation theory. Let us first consider  the theory in the absence of the defect (the bulk theory). Prior to the scaling limit in \eqref{eq:limit} the $\hat{g}$ coupling has a $\beta$ function which reads $\beta_{\hat{g}}=b\,\hat{g}^a+\cdots$ ($a=2$ in $d=4$, $a=3$ in $d=6$). Then, upon taking the limit  (\ref{eq:limit})

\begin{equation}
\beta_g=\hbar^{\frac{a}{2}\ (n-2)}\, b\,g^a+\cdots\, .
\end{equation}
Therefore, in the limit $\hbar\rightarrow 0$ with  fixed $g$, $\beta_g$ vanishes and hence the bulk theory is classical (and conformal). Note in particular that all bulk loops vanish: a diagram with $L$ bulk loops (and no interaction with the defect) is proportional to $\hat{g}^L=\hbar^{\frac{a}{2}\ L(n-2)}\,g^L$, which vanishes in this limit.

Let us now turn to the defect. One way to search for instabilities is to look for marginal/relevant operators in the defect. Such information is encoded in the correlation functions of the defect operators. Denoting generically the fields by $\Phi_i$, one would generically be interested in $\langle \Phi_{I_1}(x_1)\cdots \Phi_{I_n}(x_n)\rangle$, whose path-integral representation is

\begin{equation}
    \langle \Phi_{I_1}(x_1)\cdots \Phi_{I_r}(x_r)\rangle=\frac{1}{Z}\,\int \mathcal{D}\Phi_I \, \Phi_{I_1}(x_1)\cdots \Phi_{I_r}(x_r)\,e^{-S}\,.
\end{equation}
However, in the double-scaling limit this integral simplifies to

\begin{equation}
\label{disco}
    \langle \Phi_{I_1}(x_1)\cdots \Phi_{I_r}(x_r)\rangle=\langle \Phi_{I_1}(x_1)\rangle \cdots \langle\Phi_{I_r}(x_r)\rangle\,,
\end{equation}
where $\langle\Phi_I(x)\rangle$ is the field evaluated in the semiclassical solution obtained in section \ref{sec:Defectsinscalarfieldtheories}, which can be identified with the one-point function of $\Phi_I$. Thus, in this limit, the correlator is completely dominated by the disconnected piece.\footnote{An analogous effect occurs
in  the large $N$ limit of CFT's, where correlation functions are dominated by the disconnected term, 
due to large $N$ factorization.}
The disconnected piece is non-vanishing due to defect interactions.

The one-point function in the presence of the defect is given in eq. (4.3) in \cite{Rodriguez-Gomez:2022gif}.
This can be cast as

\begin{equation}
    \langle \Phi_I\rangle=\int \frac{d^{d_T}\vec{p}^{\, T}}{(2\pi)^{d_T}}\,\frac{e^{i\vec{p}^{\, T}\cdot\vec{x}^T}}{(\vec{p}^{\, T})^{2\,(\frac{d_T}{2}-\frac{\Delta(\Phi_I)}{2})}}\sim\frac{1}{|\vec{x}_T|^{\Delta(\Phi_I)}}\,,
    \end{equation}
    with
    \begin{equation}
    \label{deltaphis}
    \qquad \Delta(\Phi_I)=d_T-2+\frac{2\,c\,\Omega\,V_I}{u_I}-\frac{4c^3\,\Omega^2\,V_{IJ}V_J}{u_I}\,.
\end{equation}
Using this formula for the $\phi_i$ fields in the models at hand, we obtain

\begin{equation}
\label{eq:deltaperturbative}
\begin{array}{ccc}
    4d: &  \Delta(\phi_i)\Big|_{4d}=1+Q-Q^2+...\,,& Q=\frac{g\nu^2}{8\pi^2}\,;\\ &  \\ 6d: & \Delta(\phi_i)\Big|_{6d}=2+\frac{P}{2}-\frac{P^2}{8}+...\,,& P=\frac{g\nu}{2\pi^2}\,.
\end{array}
\end{equation}
{}From \eqref{disco}, it also follows  that $\Delta(\phi_{i_1}...\phi_{i_r})=r\,\Delta(\phi_i)$. 
Provided $Q,\,P\geq 0$, the $\phi_i$'s and all operators made with $\phi_i$ are irrelevant in perturbation theory, since $\Delta(\phi_i)\geq d_T-2$.
On the other hand, applying the formula \eqref{deltaphis}, one gets $\Delta(\rho)=d_T-2$, so the $\rho$ deformation is  marginal.
We shall discuss  more aspects of  stability  in the next subsection.

\subsection{Exact dimensions and instabilities}

The analysis above is just the statement that the theory indeed perturbatively defines a dCFT. However, one may fear that this is a statement only holding in perturbation theory. Indeed, in view of the dimensions above, one may worry that for large enough $Q$, $P$  one may find operators going towards marginality. Let us first consider the four-dimensional case. We note that $Q>0$ for $g>0$, that is, when the potential is positive. Thus, $\Delta(\vec{\phi})>1$ and, in perturbation theory, the theory is stable provided the potential is bounded from below. 
As a consequence, in this regime we indeed have a dCFT as anticipated, recovering exactly the same conclusions as those drawn originally from the fixed points of the $\beta$ functions. However, a question of interest is what happens for finite values of $Q$; in particular, whether instabilities as those appearing in scalar QED can appear in field theories containing only scalar fields. An exact formula for finite $Q$ in the double scaling limit can be obtained by solving saddle-point equations, which effectively
 resums the perturbative series. For the model we are studying, the equations of motion are (we now assume ``mostly minus" Minkowski signature)
 
\begin{equation}
\partial^2\phi+2\,g\rho^{2}\,\phi=0\,,\qquad \partial^2\rho + 2\,g\,\rho\,\phi^2+\nu\,\delta_T=0\,.
\end{equation}
Here we have used the $O(N)$ symmetry to align the $\vec{\phi}$ along some direction. These equations are solved by $\phi=0$ and

\begin{equation}
\rho(x)=\nu\,\int d^4y\, G(x-y)\,\delta_T(y)\,.
\end{equation}
Computing the integral one finds

\begin{equation}
\label{rhof}
\rho=\nu\, \int \frac{d^{d_T}\vec{p}_T}{(2\pi)^{d_T}} \frac{e^{i\vec{p}_T\cdot\vec{x}_T}}{(\vec{p}_T)^2}=\frac{\nu\,\Gamma(\frac{d_T}{2}-1)}{4\,\pi^{\frac{d_T}{2}}}\,\frac{1}{|\vec{x}_T|^{d_T-2}}\, ,\qquad d_T=3\ .
\end{equation}

Let us now consider time-independent fluctuations around the background. In polar coordinates as above, the eom for the $\phi $ fluctuation is 

\begin{equation}
\label{fifi}
\frac{1}{r^{2}}\,\partial_r(r^{2}\partial_r\phi)+\frac{1}{r^2}\,\Delta_{\Omega}\phi-\frac{Q}{r^2}\,\phi=0\,,
\end{equation}
where $Q$ is precisely the same $Q$ as introduced above. The general solution is given by

\begin{equation}
    \phi = \sum_{l,m}  R_{lm}(r) Y_{lm}\,, 
\end{equation}
where $Y_{lm}$ are the spherical harmonics on the $\mathbb{S}^2$, while
\begin{equation}
\label{treinta}
R_{lm}(r) = \tilde\phi_{lm}^+ \frac{1}{r^{-\frac{\Delta_l^-}{2}+1}}+\tilde\phi_{lm}^- \frac{1}{r^{-\frac{\Delta_l^+}{2}+1}}\,,
\end{equation}
with

\begin{equation}
\label{dima}
\Delta^{\pm}_l=1\pm 2\sqrt{\frac{1}{4}
+l(l+1)+Q}\,.
\end{equation}

Note that time-dependent fluctuations have the same behavior in the vicinity of $r=0$. This is seen by adding a factor $e^{iEt}$ in the ansatz for $\phi $. In  \eqref{fifi} this gives rise to a new  term $E^2\phi$, which can be neglected in the vicinity of $r=0$.

The two sets of solutions with coefficients $\tilde\phi_{lm}^+$ and $\tilde\phi_{lm}^-$ in principle define two different dCFT's according to the choice of boundary conditions. Setting $\tilde\phi_{lm}^-=0$ leaves a set of defect operators $\tilde\phi_{lm}^+$ whose  dimension can be read from \eqref{treinta} using the fact that $r$ has dimension -1 and the bulk field $\phi$ (hence $R_{lm}$) has dimension 1 in 4d
(recall that there is no bulk anomalous dimension as bulk loops are suppressed). This gives $\Delta (\tilde \phi^+_{lm} )=\frac12 \Delta^{+}_l$ or, for the alternative boundary condition, $\Delta (\tilde \phi^-_{lm} )=\frac12 \Delta^{-}_l$. 

For $g>0$, corresponding to a potential bounded from below, $Q>0$. It then follows that, since $\Delta (\tilde \phi^-_{lm} )<0$, the alternative boundary conditions are not allowed.

\smallskip

Consider now $\Delta (\tilde \phi^+_{lm} )$. It follows that $\Delta( (\tilde \phi^+_{lm})^2) = \Delta^+_l$. Expanding at small $Q$, we verify that the first corrections in $Q$ for $\Delta^{+}_{l=0}$ matches the perturbative formula \eqref{eq:deltaperturbative}.

The formula for  $\Delta^{+}_{l}$ also shows that $\Delta (\tilde \phi^+_{lm} )>1$, which implies that $\{\tilde \phi^+_{lm}\} $ correspond to irrelevant operators. General possible deformations are  composites of the schematic form $\partial_T^l\phi^k$, where $\partial_T^l$ represents the action of $l$ derivatives with respect to the transverse coordinates. They are all irrelevant operators.
Therefore in this theory there are no instabilities at any finite $Q$ and the model indeed describes a dCFT.

Instabilities appear for the theory with $g<0$, which corresponds to an unbounded potential. In this case, $Q<0$ and already  $\tilde \phi^+_{l=0}$ --corresponding to $\phi$ itself-- becomes relevant (of dimension $\Delta=\frac12 (1+\sqrt{1-|Q|})<1$),
and must be added to the defect. However, here we will not consider theories with unbounded potentials.

\medskip

Let us now comment on the 6d theory.
With no loss of generality, we may assume $g>0$, since the sign of $g$ can be flipped
by a redefinition $\rho\to-\rho$, $\nu\to-\nu$.
In this case the potential is unbounded in the negative $\rho$ direction. A similar calculation as above, leads to a background for $\rho$ given by \eqref{rhof}
with $d_T=4$. Then, one finds the following formula for the dimension of
operators $\tilde \phi^\pm_{lm}(x_{||})$ on the defect,
\begin{equation}
\label{dima6d}
\Delta (\tilde \phi^\pm_{lm} )=\frac12 \Delta^{\pm }_l\ ,\qquad \Delta^{\pm}_l=2\pm 2\sqrt{1
+l(l+2)+P}\, , \qquad P\equiv \frac{g\nu}{2\pi^2}\ .
\end{equation}
The expansion at small $P$ of the branch with + sign reproduces the perturbative result in \eqref{eq:deltaperturbative}.
We note that $P >0$ if and only if $g\nu >0$. This is precisely the case where the term of the potential $g\rho |\phi|^2$ is positive in the background provided by $\rho$, for either sign of $\nu$. In this case there is a  dCFT  defined by choosing the boundary condition $\tilde \phi^-_{lm}(x_{||})=0$. As before, the $\{ \tilde \phi^+_{lm}(x_{||})\}$ correspond to operators of the schematic form $ \partial_T^l\phi^n$, which are all irrelevant.

\subsection{A glimpse into fermion models}\label{sec:Fermions}

We now consider the possibility of constructing dCFT's involving scalar and fermion fields  using the double-scaling limit (see \cite{Herzog:2022jlx,Giombi:2022vnz} for  other interesting  studies of fermion dCFT's and boundary CFT's). This is 
feasible in four dimensions, where the Yukawa interaction is classically marginal. 

Let us consider  a Dirac fermion coupled to a real scalar field with a Yukawa interaction, with the action (using  `mostly minus' Minkowski signature)
\begin{equation}
S=\int d^4x\,\Big(i\bar{\psi}\slashed{\partial}\psi+\frac{1}{2}\left(\partial\rho \right)^2-\hat{g}\rho\,\bar{\psi}\psi+h\,\delta_T\,\rho\Big)\,.
\end{equation}
In this model the trivial line defect along $x^0$ is  deformed by a classically marginal deformation provided by the scalar itself.  
 
The double-scaling limit corresponds to consider the case 
of a large defect coupling $h$ and small bulk coupling $\hat g$;
specifically, $\hat g\to 0$ and $h\to \infty$ with $h\hat g$ fixed. This is formally implemented by the scaling  

\begin{equation}
h=\hbar^{-1}\,\nu\,,\qquad \hat{g}=\hbar\,g\,.
\end{equation}
Upon appropriately rescaling the fields, the action becomes

\begin{equation}
S=\hbar^{-1}\,\int d^4x\,\Big(i\bar{\psi}\slashed{\partial}\psi+\frac{1}{2}(\partial\rho)^2-g\rho\,\bar{\psi}\psi+\nu\,\delta_T\,\rho\Big)\,,
\end{equation}
In the $\hbar\rightarrow 0$ limit with $g$ and $\nu$ fixed, bulk loops are suppressed and quantum effects arise due to the interaction with the defect in terms of the effective coupling $g\nu $.

Let us first consider the bulk theory by itself, \textit{i.e.} let us momentarily set $\nu=0$. Prior to the scaling limit, the $\beta $  function for the $\hat{g}$ coupling is $\beta_{\hat{g}}=b\,\hat{g}^3+\cdots$. Therefore $\beta_{g}=\hbar^{2}\,b\,g^3+\cdots\rightarrow 0$ and there is no RG flow in the bulk, as expected since  bulk loops are suppressed.  The bulk theory is a CFT. We now turn to the defect. In the absence of bulk loops, no diagram can correct the $\rho$ one-point function.\footnote{Note that had we turned on a quartic bulk potential for $\rho$ this would not be the case anymore, and the defect coupling would run just.} As a result, $\nu$ does not run and the theory seems to be indeed a dCFT. This is basically a consequence of the fact that the theory contains, at least in perturbation theory, no other operator close to marginality on the line.

Besides the $\rho$ operator, the lowest scalar operator that the theory contains is $\bar{\psi}\psi$. Classically this has dimension 3, and therefore it is safely irrelevant in perturbation theory. However, it is important to understand whether for large values of the couplings this operator can hit marginality and eventually become  a relevant operator of dimension less than 1. To study the problem, we proceed as before. Like in the previous scalar model, the defect induces  a background for $\rho$ given by

\begin{equation}
    \rho=\frac{\nu}{4\pi} \,\frac{1}{|\vec x|}\ .
\end{equation}
It remains to study the fermion fluctuations in this background. Using the Dirac representation for the $\gamma$ matrices, the equation of motion for the fermion fluctuations decomposes into two equations for the Weyl spinor components $\chi$ and $\xi$

\begin{equation}
i\sigma^i\partial_i\xi-\frac{g\nu}{4\pi}\,\frac{1}{|\vec{x}|}\chi=0\,,\qquad i\sigma^i\partial_i\chi+\frac{g\nu}{4\pi}\,\frac{1}{|\vec{x}|}\xi=0\, .
\end{equation}
We are assuming time-independent fluctuations
(time dependence can be incorporated by  
a factor $e^{iEt}$, which leads to a subleading dependence near $r=0$ and it is thus unimportant for the determination of the dimension).

Solving for $\xi$ in the second equation and substituting it in the first equation, we find

\begin{equation}
\vec{\partial}^2\chi+\frac{1}{|\vec{x}|^2}\sigma^j\sigma^i\,x_j\,\partial_i\chi-\frac{Q_F}{|\vec{x}|^2}\chi=0\,,\quad Q_F=\Big(\frac{g\nu}{4\pi}\Big)^2\,.
\end{equation}
The solution to this is

\begin{equation}
\chi=\frac{1}{|\vec{x}|^{\alpha}}\,\chi_0\,,\qquad \alpha=1\pm \sqrt{1+Q_F}\,,
\end{equation}
being $\chi_0$ a constant spinor. Now, given that the dimension of a bulk fermion is $3/2$, we can write
\begin{equation}
    \Delta(\psi_0)=\frac32 -\Delta =\frac12 \mp  \sqrt{1+Q_F}
\end{equation}
In order to avoid negative dimensions, we must impose boundary conditions that keep the branch with the `+' sign. Therefore
\begin{equation}
    \Delta(\psi_0^+)=\frac12 +  \sqrt{1+Q_F}
\end{equation}
In the free-field limit, $ \Delta(\psi_0^+) \to 3/2$, as expected. It then follows that the dimension of $\bar{\psi}\psi$ is $2\Delta(\psi_0^+)$. Since $Q_F>0$ (at least for unitary theories), $\bar{\psi}\psi$ is  an irrelevant operator in the whole  region $Q_F\in (0,\infty)$
allowed by unitarity. This supports the fact that the line defect in the double-scaling limit indeed defines a dCFT. 
Finally, as a curiosity, let us comment that had we considered the case of a parity breaking theory with a potential $i\hat{g}\rho\bar{\psi}\gamma^5\psi$ we would have obtained exactly the same results. 


\section{Conclusions}\label{sec:Conclusions}

In this paper we have considered generic --that is, with an arbitrary number of scalar fields and an arbitrary marginal potential-- $d$ dimensional scalar field theories with defect deformations, in $d=4-\epsilon$ and $d=6-\epsilon$ as well as a scalar-fermion theory with a Yukawa interaction in $d=4$. All calculations are performed in a double-scaling limit \eqref{eq:limit}, where the defect couplings go to infinity and the bulk couplings go to 0. 

\medskip

We summarize the main results of this paper.

\begin{itemize}
    \item  $\beta $ functions for the defect couplings have been computed up to four loops using dimensional regularization and standard perturbation theory
    of the QFT. 

    \item The fixed points exhibit the property of {\it dimensional disentanglement}, namely the dependence on the dimension 
    appears through a universal function, which is  factorized from the coupling dependence. The universal function is the same for all fixed points, and for all models, independent of the number of scalar fields and independent of the potential. We showed that the DD property is a peculiar
    feature of the double-scaling limit and that it is not expected to hold once the full quantum effects are taken into account.

    \item DD implies that, modulo an overall scale, the  location of fixed points 
    remains unaltered as the dimension is varied. However, the RG flow depends on the dimension in a non-trivial way. In particular, an IR stable fixed point can become unstable by varying $\epsilon$ while keeping $\epsilon\ll 1$.

    \item In section \ref{sec:betas} we provide an alternative calculation of the defect $\beta $ functions 
    from the dressed Coulomb potential.
   In this scheme,  (at least up to four loops) the $\beta$ functions are  a gradient,
$\beta_i=2\,c\,\partial_i\mathcal{H}$, where $\partial_i$
    stands for derivatives with respect to the couplings $\nu_i$
    of the defect deformations $\nu_i\phi_i$.
    
  \item We have considered a few concrete examples of dCFT's and computed
  the dimension of  operators that could  lead to instabilities.
The first examples are   4d and 6d scalar field models obtained by sitting
on  particular fixed points. In all cases we showed that
all potentially dangerous operators are irrelevant
even for finite values of the coupling, therefore the dCFT's 
are stable. The results also show that instabilities may appear if one considers potentials that are unbounded from below, beyond some critical coupling.
  
\item In addition, we computed the dimension of the
bilinear fermion operator $\bar\psi\psi $ in  fermion-scalar theory with Yukawa interactions, showing that it is always irrelevant (for a real Yukawa coupling, where the theory is unitary). 

 \item In the appendix \ref{app:Rigidholography}  we provide a practical framework for rigid holography, by which one can compute $\beta $ functions to all loop orders. The  approach is essentially equivalent to the field theory calculation of section \ref{sec:betas}, being related through the conformal map. 
 The double scaling limit leads to  effects that are   analogous to the effects produced by the large $N$ limit in standard holography: it suppresses bulk loops and makes the correlation functions dominated by the disconnected term.

\item Using as an example  the scalar QED model studied in \cite{Aharony:2022ntz}, we also show that rigid holography can be used to compute the dimension
of $\bar\phi\phi $.

  \end{itemize}  
  
There remain many open questions and many interesting aspects of defect theories, which are worth of further investigation. In particular, it would be interesting to establish if  $\exp(\mathcal{H})$ is related  to the VEV of the circular defect (in the 4d theory) or to the VEV of the spherical defect (in the 6d theory) to any order in the loop expansion.  Another interesting problem is extending the application of rigid holography to theories on spaces $\mathbb{H}^{p+1}\times \mathbb{S}^{d-p-1}$ and codimension $p$ defects for other values of $p$. Other very interesting problems include understanding if dimensional disentanglement also arises in theories with fermions or vector fields, or the role of unbroken global symmetries and conformal manifolds along the lines of \cite{Drukker:2022pxk}.

\section*{Acknowledgements}

J.G.R. acknowledges financial support from projects  MINECO grant PID2019-105614GB-C21, and  from the State Agency for Research of the Spanish Ministry of Science and Inn ovation through the ``Unit of Excellence María de Maeztu 2020-2023" (CEX2019-000918-M). The work of I.C.B. and D.R.G is partly supported by Spanish national grant MCIU-22-PID2021-123021NB-I00 as well as the Principado de Asturias grant SV-PA-21-AYUD/2021/52177.

\begin{appendix}

\section{The integrals}\label{sect:integrals}

In this appendix we collect technical details of the evaluation of the integrals, borrowing results from \cite{Grozin:2003ak}. First, we compile formulas for several integrals that appear repeatedly. 

Consider the following integral
\begin{equation}
F_{x,y,z}(\vec{p}^{\, T})=\int \frac{d^{d_T}k_1}{(2\pi)^{d_T}}\,\int \frac{d^{d_T}k_2}{(2\pi)^{d_T}} \, \frac{1}{(\vec{p}^{\, T}-\vec{k}^T_1-\vec{k}^T_2)^{2x}\,(\vec{k}^T_1)^{2y}\,(\vec{k}^T_2)^{2z}} \,.
\end{equation}
Explicit evaluation gives the formula

\begin{equation}
F_{x,y,z}(\vec{p}^{\, T})=F_{x,y,z}\,(\vec{p}^{\, T})^{2\,(d_T-x-y-z)}\,,\qquad F_{x,y,z}=\frac{\pi^{d_T}}{(2\pi)^{2d_T}}\,G(z,x)\,G(y,x+z-\frac{d_T}{2})\,;
\end{equation}
being

\begin{equation}
G(n,m)\equiv \frac{\Gamma(n+m-\frac{d_T}{2})\,\Gamma(\frac{d_T}{2}-n)\,\Gamma(\frac{d_T}{2}-m)}{\Gamma(n)\,\Gamma(m)\,\Gamma(d_T-n-m)}\,.
\end{equation}

Another useful integral is

\begin{equation}
G_{x,y}(\vec{p}^{\, T})= \int \frac{d^{d_T}\vec{k}^T}{(2\pi)^{d_T}} \frac{1}{(\vec{k}^T)^{2x}\,(\vec{p}^{\, T}-\vec{k}^T)^{2y}}\,.
\end{equation}
This gives

\begin{equation}
G_{x,y}(\vec{p}^{\, T})=G_{x,y}\,|\vec{p}^{\, T}|^{2\,(\frac{d_T}{2}-x-y)}\,,\qquad G_{x,y}=\frac{\pi^{\frac{d_T}{2}}}{(2\pi)^{d_T}}\,G(x,y)\,.
\end{equation}

It will turn out to be convenient to introduce

\begin{equation}
F_x \equiv \frac{\pi^{d_T}}{(2\pi)^{2d_T}}\,G(1,x)\,G(1,\frac{2+2x-d_T}{2})\, .
\end{equation}
%

\medskip
\noindent 
Let us now  compile the results for the relevant integrals for $n=4$.

\subsubsection*{Order 0}

To order zero

\begin{equation}
\phi=\int dz_1\,G(x-z_1)\,\delta_T(z_1)= \int \frac{d^{d_T}\vec{p}}{(2\pi)^{d_T}}\, \frac{e^{i\vec{p}^{\, T}\cdot \vec{x}_T} }{|\vec{p}^{\, T}|^2}\,.
\end{equation}

\subsubsection*{Order 1}

We now need to compute $I_1$. After some manipulations 

\begin{equation}
I_1=\int \frac{d^{d_T}p}{(2\pi)^d}\,\frac{e^{i\vec{p}^{\, T}\cdot\vec{x}^T}}{(\vec{p}^{\, T})^2}\, \mathcal{I}_1\,,\qquad \mathcal{I}_1=\int \frac{d^{d_T}p_1}{(2\pi)^d}\,\int \frac{d^{d_T}p_2}{(2\pi)^d} \, \frac{1}{(\vec{p}^T-\vec{p}^T_1-\vec{p}^T_2)^2\,(\vec{p}^T_1)^2\,(\vec{p}^T_2)^2} \,.
\end{equation}
Using the formulas above, we see that

\begin{equation}
\mathcal{I}_1=F_{1,1,1}\,|\vec{p}^{\, T}|^{2(d_T-3)}\,.
\end{equation}

\subsubsection*{Order 2}

We now need $I_2$, which can be re-written as

\begin{equation}
I_2=\int \frac{d^{d_T}\vec{p}^{\, T}}{(2\pi)^{d_T}}\,\frac{e^{i\vec{p}^{\, T}\cdot\vec{x}^T}}{(\vec{p}^{\, T})^2}\,\mathcal{I}_2\,,
\end{equation}
with
\begin{eqnarray}
\mathcal{I}_2=&&\int \frac{d^{d_T}\vec{k}^T_3}{(2\pi)^{d_T}}\,\frac{1}{(\vec{k}^T_3)^2}\\ \nonumber && \Big[\int \frac{d^{d_T}\vec{k}^T_1}{(2\pi)^{d_T}}\int \frac{d^{d_T}\vec{k}^T_2}{(2\pi)^{d_T}} \frac{1}{(\vec{k}^T_1)^2\,(\vec{k}^T_2)^2\,(\vec{k}^T_3-\vec{k}^T_1-\vec{k}^T_2)^2}\Big] \, \Big[ \int \frac{d^{d_T}\vec{k}^T_4}{(2\pi)^{d_T}} \frac{1}{(\vec{k}^T_4)^2\,(\vec{p}^{\, T}-\vec{k}^T_3-\vec{k}^T_4)^2} \Big]\,.
\end{eqnarray}

Using the results for the integrals above

\begin{equation}
\mathcal{I}_2=F_{1,1,1}\,G_{1,1}\,G_{4-d_T,\frac{4-d_T}{2}}\,|\vec{p}^{\, T}|^{2\,(2d_T-6)}\,\,.
\end{equation}

\subsubsection*{Order 3}

Now we have two integrals

\begin{itemize}

\item $\mathbf{I_3^{(1)}}$: after some tedious but straightforward manipulations, one can show that 

\begin{equation}
I_3^{(1)}=\int \frac{d^{d_T}p}{(2\pi)^{d_T}}\frac{e^{i\vec{p}^{\, T}\cdot\vec{x}^T}}{(\vec{p}^{\, T})^2}\, \mathcal{I}_3^{(1)}\,,
\end{equation}
where

\begin{eqnarray}
 \mathcal{I}_3^{(1)}&=&\int \frac{d^{d_T}k_1}{(2\pi)^{d_T}}\int \frac{d^{d_T}k_2}{(2\pi)^{d_T}}\frac{1}{(\vec{k}_1^T)^2\,(\vec{k}_2^T)^2\,(\vec{p}^{\, T}-\vec{k}_1^T-\vec{k}_2^T)^2}\\ \nonumber  && \int \frac{d^{d_T}k_3}{(2\pi)^{d_T}}\int \frac{d^{d_T}k_4}{(2\pi)^{d_T}}\frac{1}{(\vec{k}_4^T)^2\,(\vec{k}_3^T)^2\,(\vec{p}^{\, T}-\vec{k}_1^T-\vec{k}_2^T-\vec{k}_3^T-\vec{k}_4^T)^2}\\ \nonumber  && \Big[\int \frac{d^{d_T}k_5}{(2\pi)^{d_T}}\int \frac{d^{d_T}k_6}{(2\pi)^{d_T}}\frac{1}{(\vec{k}_5^T)^2\,(\vec{k}_6^T)^2\,(\vec{p}^{\, T}-\vec{k}_1^T-\vec{k}_2^T-\vec{k}_3^T-\vec{k}_4^T-\vec{k}_5^T-\vec{k}_6^T)^2}\Big]\,.
\end{eqnarray}

Using the results above

\begin{eqnarray}
&& \mathcal{I}_3^{(1)}=F_{1,1,1}\,F_{4-d_T,1,1}\,F_{7-2d_T,1,1} |\vec{p}^{\, T}|^{2\,(3d_T-9)}\,.
\end{eqnarray}
This can be written as
\begin{eqnarray}
&& \mathcal{I}_3^{(1)}=F_1\,F_{4-d_T}\,F_{7-2d_T} |\vec{p}^{\, T}|^{2\,(3d_T-9)}\,.
\end{eqnarray}

\item $\mathbf{I_3^{(2)}}$: in this case, one finds
\begin{equation}
I_3^{(2)}=\int \frac{d^{d_T}p}{(2\pi)^{d_T}}\frac{e^{i\vec{p}^{\, T}\cdot\vec{x}^T}}{(\vec{p}^{\, T})^2}\, \mathcal{I}_3^{(2)}\,,
\end{equation}
where
\begin{eqnarray}
 \mathcal{I}_3^{(2)}&=& \int \frac{d^{d_T}p_1}{(2\pi)^{d_T}}\int \frac{d^{d_T}p_2}{(2\pi)^{d_T}} \frac{1}{(\vec{p}_1^T)^2\,(\vec{p}_2^T)^2\,(\vec{p}-\vec{p}_1-\vec{p}_2)^2} \\ \nonumber &&\Big[\int \frac{d^{d_T}k_1}{(2\pi)^{d_T}}\int \frac{d^{d_T}k_2}{(2\pi)^{d_T}}\frac{1}{(\vec{k}_1^T)^2\,(\vec{k}_2^T)^2\,(\vec{p}_1^T-\vec{k}_1^T-\vec{k}_2^T)^2}\Big]\\ \nonumber &&  \Big[\int \frac{d^{d_T}k_3}{(2\pi)^{d_T}}\int \frac{d^{d_T}k_4}{(2\pi)^{d_T}}\frac{1}{(\vec{k}_3^T)^2\,(\vec{k}_4^T)^2\,(\vec{p}_2^T-\vec{k}_3^T-\vec{k}_4^T)^2}\Big]\,.
\end{eqnarray}

Using the results above

\begin{equation}
 \mathcal{I}_3^{(2)}= F_{1,1,1}^2\, F_{1,4-d_T,4-d_T}\,|\vec{p}^{\, T}|^{2(3d_T-9)} \,.
\end{equation}
This can be rewritten as
\begin{equation}
 \mathcal{I}_3^{(2)}= F_1\,F_{4-d_T}\,F_{7-2d_T}\,|\vec{p}^{\, T}|^{2(3d_T-9)}\,\frac{2\,(1-5\epsilon)}{1-3\epsilon}\ \,.
\end{equation}
It then follows that
\begin{equation}
 \mathcal{I}_3^{(2)}=\frac{2\,(1-5\epsilon)}{1-3\epsilon}\, \mathcal{I}_3^{(1)} \qquad \implies \qquad
 I_3^{(2)}=\frac{2\,(1-5\epsilon)}{1-3\epsilon}\, I_3^{(1)} \,.
\end{equation}

\end{itemize}

\subsubsection*{Order 4}

Now we have 4 integrals

\begin{itemize}
\item $\mathbf{I_4^{(1)}}$: we have
\begin{equation}
I_4^{(1)}=\int \frac{d^{d_T}\vec{p}^{\, T}}{(2\pi)^{d_T}}\,\frac{e^{i\vec{p}^{\, T}\cdot\vec{x}^T}}{(\vec{p}^{\, T})^2}\,\mathcal{I}_4^{(1)}\,,
\end{equation}
where
\begin{eqnarray}
 \mathcal{I}_4^{(1)}&=& \int \frac{d^{d_T}k_1}{(2\pi)^{d_T}}\,\int \frac{d^{d_T}k_2}{(2\pi)^{d_T}}\,\frac{1}{(\vec{k}_1^T)^2\,(\vec{k}^T_2)^2\,(\vec{p}-\vec{k}_1^T-\vec{k}_2^T)^2} 
\\ \nonumber && \int \frac{d^{d_T}k_3}{(2\pi)^{d_T}}\,\int \frac{d^{d_T}k_4}{(2\pi)^{d_T}}\,\frac{1}{(\vec{k}_3^T)^2\,(\vec{k}^T_4)^2\,(\vec{p}-\vec{k}_1^T-\vec{k}_2^T-\vec{k}_3^T-\vec{k}_4^T)^2} 
\\ \nonumber && \int \frac{d^{d_T}k_5}{(2\pi)^{d_T}}\,\int \frac{d^{d_T}k_6}{(2\pi)^{d_T}}\,\frac{1}{(\vec{k}_5^T)^2\,(\vec{k}^T_6)^2\,(\vec{p}-\vec{k}_1^T-\vec{k}_2^T-\vec{k}_3^T-\vec{k}_4^T-\vec{k}_5^T-\vec{k}_6^T)^2} 
\\ \nonumber && \Big[\int \frac{d^{d_T}k_7}{(2\pi)^{d_T}}\, \int \frac{d^{d_T}k_8}{(2\pi)^{d_T}} \frac{1}{(\vec{k}_7^T)^2\,(\vec{k}^T_8)^2\,(\vec{p}^{\, T}-\vec{k}_1^T-\vec{k}_2^T-\vec{k}_3^T-\vec{k}_4^T-\vec{k}_5^T-\vec{k}_6^T-\vec{k}_7^T-\vec{k}_8^T)^2}\Big]
\end{eqnarray}

Using the formulas above
\begin{equation}
\mathcal{I}_4^{(1)}=F_{1,1,1}\,F_{4-d_T,1,1}\,F_{7-2d_T,1,1}\,F_{10-3d_T,1,1}\,(\vec{p})^{2\,(4d_T-12)}\,.
\end{equation}

This is nicest rewritten as follows:
\begin{equation}
\mathcal{I}_4^{(1)}=F_1\,F_{4-dT}\,F_{7-2d_T}\,F_{10-3d_T}\,(\vec{p})^{2\,(4d_T-12)}\,.
\end{equation}

\item $\mathbf{I_4^{(2)}}$: we have
\begin{equation}
I_4^{(2)}=\int \frac{d^{d_T}\vec{p}^{\, T}}{(2\pi)^{d_T}}\,\frac{e^{i\vec{p}^{\, T}\cdot\vec{x}^T}}{(\vec{p}^{\, T})^2}\,\mathcal{I}_4^{(2)}\,,
\end{equation}
where
 \begin{eqnarray}
 \mathcal{I}_4^{(2)}&=& \int \frac{d^{d_T}k_1}{(2\pi)^{d_T}}\, \int \frac{d^{d_T}k_2}{(2\pi)^{d_T}}\, \frac{1}{(\vec{k}_1^T)^2\,(\vec{k}_2^T)^2\,(\vec{p}^{\, T}-\vec{k}_1^T-\vec{k}_2^T)^2}\\ \nonumber &&  \int \frac{d^{d_T}k_3}{(2\pi)^{d_T}}\, \int \frac{d^{d_T}k_4}{(2\pi)^{d_T}}\, \frac{1}{(\vec{k}_3^T)^2\,(\vec{k}_4^T)^2\,(\vec{p}^{\, T}-\vec{k}_1^T-\vec{k}_2^T-\vec{k}_3^T-\vec{k}_4^T)^2} \\ \nonumber && \Big[ \int \frac{d^{d_T}k_5}{(2\pi)^{d_T}}\, \int \frac{d^{d_T}k_6}{(2\pi)^{d_T}}\,\frac{1}{(\vec{k}_5^T)^2\,(\vec{k}_6^T)^2\,(\vec{k}_3^T-\vec{k}_5^T-\vec{k}_6^T)^2}\Big] \\ \nonumber &&\Big[ \int \frac{d^{d_T}k_7}{(2\pi)^{d_T}}\, \int \frac{d^{d_T}k_8}{(2\pi)^{d_T}}\,\frac{1}{(\vec{k}_7^T)^2\,(\vec{k}_8^T)^2\,(\vec{k}_4^T-\vec{k}_7^T-\vec{k}_8^T)^2}\Big]\,.
\end{eqnarray}

Using the results above

\begin{equation}
\mathcal{I}_4^{(2)}=F_{1,1,1}^2\,F_{1,4-d_T,4-d_T}\,F_{10-3d_T,1,1}\,(\vec{p})^{2\,(4d_T-12)}\,.
\end{equation}
One can check that this can be rewritten as

\begin{equation}
\mathcal{I}_4^{(2)}=\frac{2\,(1-5\epsilon)}{1-3\epsilon}\,\mathcal{I}_4^{(1)}\,.
\end{equation}

\item $\mathbf{I_4^{(3)}}$

We have

\begin{equation}
I_4^{(2)}=\int \frac{d^{d_T}\vec{p}^{\, T}}{(2\pi)^{d_T}}\,\frac{e^{i\vec{p}^{\, T}\cdot\vec{x}^T}}{(\vec{p}^{\, T})^2}\,\mathcal{I}_4^{(3)}\,,
\end{equation}
where

 \begin{eqnarray}
&& \mathcal{I}_4^{(3)}=\int \frac{d^{d_T}k_1}{(2\pi)^{d_T}}\, \int \frac{d^{d_T}k_2}{(2\pi)^{d_T}}\,  \frac{1}{(\vec{k}_1^T)^2\,(\vec{k}_2^T)^2\,(\vec{p}^{\, T}-\vec{k}_1^T-\vec{k}_2^T)^2} \\ \nonumber && \Big[ \int \frac{d^{d_T}k_3}{(2\pi)^{d_T}}\, \int \frac{d^{d_T}k_4}{(2\pi)^{d_T}}\, \frac{1}{(\vec{k}_3^T)^2\,(\vec{k}_4^T)^2\,(\vec{k}_2^T-\vec{k}_3^T-\vec{k}_4^T)^2} \Big] \\ \nonumber  &&\int \frac{d^{d_T}k_5}{(2\pi)^{d_T}}\, \int \frac{d^{d_T}k_6}{(2\pi)^{d_T}}\,\frac{1}{(\vec{k}_5^T)^2\,(\vec{k}_6^T)^2\,(\vec{k}_1^T-\vec{k}_5^T-\vec{k}_6^T)^2} \\ \nonumber &&\Big[ \int \frac{d^{d_T}k_7}{(2\pi)^{d_T}}\, \int \frac{d^{d_T}k_8}{(2\pi)^{d_T}}\,\frac{1}{(\vec{k}_7^T)^2\,(\vec{k}_8^T)^2\,(\vec{k}_5^T-\vec{k}_7^T-\vec{k}_8^T)^2}\Big]
\end{eqnarray}

Using the integrals above
\begin{equation}
\mathcal{I}_4^{(3)}=F_{1,1,1}^2\,F_{1,4-d_T,1}\,F_{1,7-2d_T,4-d_T}\,\,(\vec{p})^{2\,(4d_T-12)}\, ,
\end{equation}
which can be rewritten as
\begin{equation}
\mathcal{I}_4^{(3)}=\frac{3\,(1-7\epsilon)}{1-3\epsilon}\,\mathcal{I}_4^{(1)}\,.
\end{equation}

\item $\mathbf{I_4^{(4)}}$: we have
\begin{equation}
I_4^{(2)}=\int \frac{d^{d_T}\vec{p}^{\, T}}{(2\pi)^{d_T}}\,\frac{e^{i\vec{p}^{\, T}\cdot\vec{x}^T}}{(\vec{p}^{\, T})^2}\,\mathcal{I}_4^{(4)}\,,
\end{equation}
where
 \begin{eqnarray}
&& \mathcal{I}_4^{(4)}=\int \frac{d^{d_T}k_7}{(2\pi)^{d_T}}\, \int \frac{d^{d_T}k_8}{(2\pi)^{d_T}}\, \frac{1}{(\vec{k}_7^T)^2\,(\vec{k}_8^T)^2\,(\vec{p}^{\, T}-\vec{k}_7-\vec{k}_8^T)^2}  \\ \nonumber && \Big[\int \frac{d^{d_T}k_5}{(2\pi)^{d_T}}\, \int \frac{d^{d_T}k_6}{(2\pi)^{d_T}}\, \frac{1}{(\vec{k}_5^T)^2\,(\vec{k}_6^T)^2\,(\vec{k}_8^T-\vec{k}_5^T-\vec{k}_6^T)^2} \Big] \\ \nonumber && \Big[\int \frac{d^{d_T}k_3}{(2\pi)^{d_T}}\, \int \frac{d^{d_T}k_4}{(2\pi)^{d_T}}\, \frac{1}{(\vec{k}_3^T)^2\,(\vec{k}_4^T)^2\,(\vec{p}^{\, T}-\vec{k}_7^T-\vec{k}_8^T-\vec{k}_3^T-\vec{k}_4^T)^2}\Big] \\ \nonumber && \Big[\int \frac{d^{d_T}k_1}{(2\pi)^{d_T}}\, \int \frac{d^{d_T}k_2}{(2\pi)^{d_T}}\, \frac{1}{(\vec{k}_1^T)^2\,(\vec{k}_2^T)^2\,(\vec{k}_7^T-\vec{k}_1^T-\vec{k}_2^T)^2}\Big]
\end{eqnarray}

Using the formulas above
\begin{equation}
\mathcal{I}_4^{(4)}=F_{1,1,1}^3\,F_{4-d_T,4-d_T,4-d_T}\,\,(\vec{p})^{2\,(4d_T-12)}\, ,
\end{equation}
which reduces to the simpler form
\begin{equation}
\mathcal{I}_4^{(4)}=\frac{6\,(1-5\epsilon)(1-7\epsilon)}{(1-3\epsilon)^2}\,\mathcal{I}_4^{(1)}\,.
\end{equation}

\end{itemize}

\section{Four-loop $\beta$ functions }\label{sect:holo4}

In this appendix we calculate the $\beta$ function  to four-loop order using the approach of section 3.
Our starting point is the equation of motion of $f_i^{(4)}$  as defined in  \eqref{eq:eomradial}. This is
\begin{equation}
    \partial_r\left({r^{3-d_T}}\partial_r f_i^{(4)}\right)-\frac{1}{r^{d_T-1}}\,\Big\{ V_{ij} f_j^{(3)}+ V_{ijk} f_j^{(1)} f_k^{(2)}+\frac16 V_ {ijkl}f_j^{(1)}f_k^{(1)}f_l^{(1)}\Big\}=0\,,
\nonumber
\end{equation}
where the $f_i^{(n)}$ are defined at \eqref{eq:fpert}.
 The solution is given by

\begin{equation}
\begin{aligned}
    f_i^{(4)}&=\frac{ V_{ij} V_{jk} V_{kl} V_l }{24 p^7} \left(p^3 (\log r)^4+12 p^2 (\log r)^3+60
   p (\log r)^2+120 \log r\right)\\
   &+\frac{  V_{ij} V_{jkl} V_k V_l}{24 p^7} \left(p^3 (\log r)^4+8 p^2 (\log r)^3+36 p (\log r)^2+72 \log r\right)\\
   &+\frac{ V_{ijk} V_j V_{kl} V_l }{24 p^7} \left(3 p^3 (\log r)^4+20 p^2 (\log r)^3+60 p (\log r)^2+120 \log r\right)\\
   &+\frac{ V_{ijkl} V_j V_k V_l}{24 p^7}\left(p^3 (\log r)^4+4 p^2 (\log r)^3+12 p (\log r)^2+24 \log r\right)\, .
   \end{aligned}
\end{equation}

where $p\equiv d_T-2$, {\it i.e.} $p=1$ in 4d and $p=2$ in $6d$. With this result, following the same steps as in section \ref{sec:betas}, we define a ``running'' $h_i$ as
\begin{equation}
\begin{aligned}
u_i&=h_i-\frac{2\Omega\, V_i}{p}\,\log r+\frac{2\Omega^2 V_{ij}V_j}{p^3}\,\big(2\log r+p\,(\log r)^2\big)  \\
&- \frac{4\Omega^3 V_{ij} V_{jk} V_k}{3 p^5 }\left (  p^2 (\log r)^3+6  p (\log r)^2 +12 \log r\right)\\
&- \frac{4 \Omega^3 V_{ijk} V_j V_k}{3 p^5 } \left(  p^2 (\log r)^3+3 p (\log r)^2+6 \log r \right)\\
&+\frac{2 \Omega^4 V_{ij} V_{jk} V_{kl} V_l }{3 p^7} \left(p^3 (\log r)^4+12 p^2 (\log r)^3+60
   p (\log r)^2+120 \log r\right)\\
   &+\frac{2 \Omega^4  V_{ij} V_{jkl} V_k V_l}{3 p^7} \left(p^3 (\log r)^4+8 p^2 (\log r)^3+36 p (\log r)^2+72 \log r\right)\\
   &+\frac{2 \Omega^4 V_{ijk} V_j V_{kl} V_l }{3 p^7} \left(3 p^3 (\log r)^4+20 p^2 (\log r)^3+60 p (\log r)^2+120 \log r\right)\\
   &+\frac{2 \Omega^4 V_{ijkl} V_j V_k V_l}{3 p^7}\left(p^3 (\log r)^4+4 p^2 (\log r)^3+12 p (\log r)^2+24 \log r\right)\, ,
\end{aligned}
\end{equation}
where the $V$ are evaluated at $h_i$. Inverting this formula we get
\begin{equation}
\begin{aligned}
h_i&=u_i+\frac{2\Omega\, V_i}{p}\,\log r-\frac{2\Omega^2 V_{ij}V_j}{p^3}\,\big(2\log r-p\,(\log r)^2\big) \\
&+\frac{4\Omega^3 V_{ij} V_{jk} V_k}{3 p^5}\left(p^2 (\log r)^3-6 p (\log r)^2+12 \log r\right)\\
&+\frac{4\Omega^3V_{ijk} V_{j} V_k}{3 p^5} \left( p^2 (\log r)^3-3 p (\log r)^2+6 \log r \right)\\
&-\frac{2 \Omega ^4  V_{ij} V_{jk} V_{kl} V_l}{3 p^7}\left(- p^3 (\log r)^4+12
   p^2 (\log r)^3-60 p (\log r)^2+120\log r\right) \\
   &-    \frac{2 \Omega ^4 V_{ij} V_{jkl} V_k V_l }{3
   p^7}\left(11 p^3 (\log r)^4+8
   p^3 (\log r)^3-36 p (\log r)^2+72 \log r\right)\\  
   &- \frac{2 \Omega ^4  V_{ijk} V_j V_{kl} V_l  }{3
   p^7} \left(-15 p^3 (\log r)^4+20
   p^2 (\log r)^3-60 p (\log r)^2+120\log r\right)
   \\
   &-  \frac{2 \Omega ^4 V_{ijkl} V_j V_k V_l  }{3
   p^7} \left(- p^3 (\log r)^4+4
   p^2 (\log r)^3-12 p (\log r)^2+24 \log r\right)\,.
\end{aligned}
\end{equation}
Now we interpret once again $r^{-1}$ as the RG scale and differentiate both sides of the equation with respect to $\log r$ to get the $\beta$ function for $u_i$. We obtain
\begin{equation}
\begin{aligned}
\label{eq:betainholographyo4}
\beta_i&=2\,c\,\Omega\,V_i-4\,c^3\,\Omega^2\,V_{ij}V_j+8\,c^5\,\Omega^3\,(V_{ijk}V_j\,V_k+2\,V_{ij}V_{jk}\,V_k)-\\
&-80\, c^7 \,\Omega^4 \left(\,V_{ij} V_{jk} V_{kl} V_l+ \,V_{ijk} V_{jk} V_{k} V_l\right)-16\, c^7 \,\Omega^4\left( V_{ijkl} V_j V_k V_l+3\,  V_{ij} V_{jkl} V_j V_l\right) \, , 
\end{aligned}
\end{equation}
where $c=1/p=(d_T-2)^{-1}$. From this formula, we compute the function $\mathcal{H}$ found in \eqref{eq:hgrad} up to fourth loop order. 


\section{Rigid holography}\label{app:Rigidholography}

In this section we will show that the $\beta $ functions and dimensions can also be computed  using holographic techniques. As it is well known, $\mathbb{R}^d$ can be conformally mapped to $\mathbb{H}^{a+1}\times \mathbb{S}^{b+1}$ with $a+b+2=d$. To see this, we start with the $\mathbb{R}^d$ metric, written as

\begin{equation}
ds^2_{\mathbb{R}^d}=dr_1^2+r_1^2\,ds_{\mathbb{S}^a}^2+dr_2^2+r_2^2\,ds_{\mathbb{S}^b}^2\,,\qquad a+b+2=d\,.
\end{equation}
Next,  we  perform the following change of coordinates

\begin{equation}
r_1=\frac{\sinh\rho}{\cosh\rho-\cos\psi}\,,\qquad r_2=\frac{\sin\psi}{\cosh\rho-\cos\psi}\,.
\end{equation}
Then, the metric becomes

\begin{equation}
ds_{\mathbb{R}^d}=F^2\,\Big( d\rho^2+\sinh^2\rho\,ds_{\mathbb{S}^a}^2+d\psi^2+\sin^2\psi\,ds^2_{\mathbb{S}^b} \Big)\,,\qquad F=\frac{1}{\cosh\rho-\cos\psi}\,.
\end{equation}
Up to a conformal factor, this is $\mathbb{H}^{a+1}\times \mathbb{S}^{b+1}$. Note that the boundary sits at $\rho\rightarrow \infty$, which corresponds to $r_1=1$, $r_2=0$. Thus, the boundary of $\mathbb{H}^{a+1}$ is conformal to the $\mathbb{S}^a$ at $r_1=1$ and $r_2=0$ in the original coordinates, which is in turn conformal to $\mathbb{R}^a$. In fact, we can directly have such flat boundary by considering the hyperbolic space in Poincar\'e coordinates.

\subsection{$\beta$ functions from rigid holography}

Since in the double scaling limit our theories are conformal --at least in the bulk-- we can make use of this conformal transformation to have the defect living at the boundary of the Poincar\'e hyperbolic space.\footnote{The double scaling limit plays an analogous role to the large $N$ limit in standard holography; in both cases the limit leads to a classical bulk theory.} 
In the cases of interest we need $a=b=d_T-2$, corresponding to $\mathbb{H}^{d_T-1}\times \mathbb{S}^{d_T-1}$, such that in $d=4$, where $d_T=3$, we have $\mathbb{H}^2\times \mathbb{S}^2$;  while in $d=6$ (where $d_T=4$) we have $\mathbb{H}^3\times \mathbb{S}^3$. The metric of the $\mathbb{H}^{d_T-1}\times \mathbb{S}^{d_T-1}$ space is
\begin{equation}
\label{metricaads}
ds^2=\frac{dz^2+ds_{\mathbb{R}^{d_T-2}}^2}{z^2}+ds_{\mathbb{S}^{d_T-1}}^2\,.
\end{equation}

Since in  $\mathbb{H}^{d_T-1}\times \mathbb{S}^{d_T-1}$ the conformal coupling to curvature for scalars is zero, the bulk action lagrangian is simply the original one in the curved  $\mathbb{H}^{d_T-1}\times \mathbb{S}^{d_T-1}$ space with metric \eqref{metricaads}
\begin{equation}
S=\int d^dx \left(\frac{1}{2}(\partial\phi_i)^2+V(\phi_i)\right)\,.
\end{equation}
All in all the problem is mapped to a holographic scenario albeit without gravity. This is very reminiscent of the \textit{rigid holography} scenario of \cite{Aharony:2015zea} (for further developments along these lines, see \textit{e.g.} \cite{Paulos:2016fap,Beccaria:2017rbe,Carmi:2018qzm,Herzog:2019bom,Herzog:2020lel,Giombi:2020amn,Giombi:2020rmc,Giombi:2021cnr}).

For $z$-dependent configurations, the equations of motion are
\begin{equation}
\partial_z\big(\frac{1}{z^{d_T-1}}z^2\partial_z\phi_i\big)-\frac{1}{z^{d_T-1}}\,V_i=0\,,
\end{equation}
where $V_i=\frac{\partial V}{\partial\phi_i}$. This equation is identical to \eqref{eq:eomradial} upon changing $z\rightarrow r$. As a consequence, the solution can be immediately borrowed from there
\begin{equation}
    \phi_i=u_i\,,\qquad u_i=s_i+f^{(1)}_i+f^{(2)}_i+\cdots\,,
\end{equation}
with the $f_i^{(k)}$ being the same as in \eqref{eq:fpert} upon changing $r$ by $z$. The rest of the computation goes by unchanged, leading to exactly the same $\beta$-functions. 

The appearance of the $\beta $ functions through rigid holography is reminiscent of the Wilson loop case \cite{Alday:2007he,Polchinski:2011im,Beccaria:2017rbe}, where there is a flow between the Wilson loop and the Wilson-Maldacena loop (see also \cite{Drukker:2020swu} for the membrane case). It is interesting to observe that the double-scaling limit has an effect similar to the large $N$ limit in standard holography (freezing bulk loops and leading to ``large $N$ factorization").
 
\subsection{Dimension of gauge-invariant operators in scalar QED}\label{subsec:holoscalarqed}

In a recent paper \cite{Aharony:2022ntz}, Aharony {\it et al.} studied phase transitions in scalar  QED with a Wilson line, by computing the dimension of scalar operators on the defect.
Rigid holography can also be used to reproduce the results in \cite{Aharony:2022ntz}. We consider a Wilson line along, say, $x^0$ in $\mathbb{R}^{1,3}$. Assuming mostly minus signature, the action is (we follow the conventions in \cite{Aharony:2022ntz})

\begin{equation}
    S=\int d^4x\, \sqrt{-g}\left(-\frac{1}{4e^2}F^2+|\partial\phi-i e A \phi|^2-\frac{\hat{\lambda}}{2}|\phi|^4-\hat{q}\,A_0\,\delta_T \right)\,.
\end{equation}
Introducing now $\hat{\lambda}=\lambda\,e^2$ and $\hat{q}=q\,e^{-2}$ and appropriately rescaling the fields, we can write
\begin{equation}
    S=\frac{1}{e^2}\,S_{\rm eff}\,,\qquad S_{\rm eff}=\int d^4x\, \sqrt{-g}\left(-\frac{1}{4}F^2+|\partial\phi-i e\,A \phi|^2-\frac{\lambda}{2}|\phi|^4-q\,A_0\,\delta_T\right)\,.
\end{equation}
We now take the semiclassical limit $e\rightarrow 0$ with $q$ and $\lambda$ fixed. In this limit bulk loops are suppressed. Since $q$ cannot run due to gauge invariance, we again find, naively, a dCFT.

Let us now map the problem to $AdS_2\times \mathbb{S}^2$, with metric
\eqref{metricaads}. Assuming an ansatz $A_0=A_0(z)$ and $\phi=\phi(z)$, the bulk equations of motion become 

\begin{equation}
    \partial_z(z^2\partial_zA_0)+2\,e^2\,A_0\,|\phi|^2=0\,,\qquad \partial_z^2\phi+e^2\,A_0^2\,\phi-z^{-2}\,\lambda\,|\phi|^2\,\phi=0\,.
\end{equation}
Let us now look for the appropriate holographic configuration representing charge source.
The general background solution for $A_0$ with $z$-dependence 
and $\phi=0$ is given by
$$
A_0=a_0+j_0\,z^{-1}\ .
$$
 We will choose the boundary condition $a_0=0$, which corresponds to a current --as opposed to a dynamical gauge field-- in the boundary \cite{Marolf:2006nd}. Moreover, just as in the scalar case, the $j_0$ constant is fixed by the Coulomb law as
\begin{equation}
    j_0=-\frac{q}{4\pi}\,.
\end{equation}

Turning now to the equation for the $\phi$ fluctuations in this background, to quadratic order one finds
\begin{equation}
\partial_z^2\phi+\frac{Q}{4\,z^2}\,\phi=0\,,\qquad Q=\frac{e^2\,q^2}{4\pi^2}\ .
\end{equation}
The solution to this equation is
\begin{equation}
    \phi=C_+\,z^{\frac{\Delta_+}{2}}+C_-\,z^{\frac{\Delta_-}{2}}\,,\qquad \Delta_{\pm}=1\pm \sqrt{1-Q}\,.
\end{equation}
Note that in terms of the original variables
\begin{equation}
    Q=\frac{e^4\,\hat{q}^2}{4\pi^2}\, .
\end{equation}
This reproduces the results in \cite{Aharony:2022ntz}, now by using holography.

As discussed in \cite{Aharony:2022ntz}, there are two possible quantizations, corresponding  to the two possible boundary
conditions $C_+=0$ or $C_-=0$. In one quantization, the Wilson line defines a stable dCFT with $|\phi|^2$ being an irrelevant deformation of dimension $\Delta_+=1+\sqrt{1-Q}$. The other
quantization defines an unstable dCFT where $|\phi|^2$ is a relevant deformation of dimension $\Delta_-=1-\sqrt{1-Q}$. As $Q$ is increased, at $Q=1$ these two branches approach and merge, resulting in fixed point annihilation and conformality loss. From the viewpoint of the stable dCFT, the naively irrelevant operator $|\phi|^2$  decreases its dimension and eventually becomes marginal at $Q=1$ (in standard terminology, it is a dangerously irrelevant operator).

It is instructive to compare  with the scalar field models, where instabilities only appeared for potentials with the wrong sign. This is consistent with the fact that in scalar QED the term $A_0^2\bar\phi\phi $ contributes with negative sign to the effective potential.
As a result, the effective charge $Q$ appears with negative sign inside the square root, leading to instabilities at critical values.

\end{appendix}

\end{document}